\documentclass[3p,final,times,sort&compress]{elsarticle}
\usepackage[utf8]{inputenc}
\usepackage[T1]{fontenc}
\usepackage{bm}
\usepackage{graphicx}
\usepackage{amsmath}
\usepackage{titlesec}
\usepackage{algorithm}
\usepackage{algpseudocode}
\usepackage{bbold}
\usepackage{mathtools}
\usepackage[font={small,it}]{caption}
\usepackage[superscript,biblabel]{cite}
\usepackage[parfill]{parskip}

\newcommand{\bold}[1]{\mathbf{#1}}

\newcommand{\postscript}[2]{{{#2}^{#1}}}

\newcommand{\dd}[0]{\text{ }\mathrm{d}}

\journal{Computer Methods in Applied Mechanics and Engineering}

\begin{document}

\title{A Fluid-Solid-Growth Solver for Cardiovascular Modeling}

\author[inst3]{Erica~L. Schwarz}
\author[inst1]{Martin~R. Pfaller}
\author[inst1]{Jason~M. Szafron}
\author[inst2]{Marcos Latorre}
\author[inst1]{Stephanie~E. Lindsey}
\author[inst5]{Christopher~K. Breuer}
\author[inst4]{Jay~D. Humphrey}
\author[inst1]{Alison~L. Marsden}

\affiliation[inst3]{
organization={Department of Bioengineering, Stanford Univeristy},
city={Stanford},
state={CA 94306},
country={USA}}

\affiliation[inst1]{
organization={Department of Pediatrics -- Cardiology, Stanford Univeristy},
city={Stanford},
state={CA 94306},
country={USA}}

\affiliation[inst2]{
organization={Center for Research and Innovation in Bioengineering, Universitat Politècnica de València},
city={València 46022},
country={Spain}}

\affiliation[inst5]{
organization={Department of Surgery, The Ohio State University Wexner Medical Center},
city={Columbus},
state={OH 43210},
country={USA}}

\affiliation[inst4]{
organization={Department of Biomedical Engineering, Yale Univeristy},
city={New Haven},
state={CT 06520},
country={USA}}

\begin{abstract}

We implement full, three-dimensional constrained mixture theory for vascular growth and remodeling into a finite element fluid-structure interaction (FSI) solver. The resulting "fluid-solid-growth" (FSG) solver allows long term, patient-specific predictions of changing hemodynamics, vessel wall morphology, tissue composition, and material properties. This extension from short term (FSI) to long term (FSG) simulations increases clinical relevance by enabling mechanobioloigcally-dependent studies of disease progression in complex domains.

\end{abstract}

\maketitle

\section{Introduction}

It is well known that blood vessels are highly sensitive to perturbations in their mechanical environment. Early work proposed phenomenological mechanoregulated frameworks for growth and remodeling (G\&R) that restored models of vascular tissues to a homeostatic mechanobiological state.\cite{rodriguez1994stress, fung1993remodeling} Since then, increasingly mechanistic models of vessel evolution have been developed, including the constrained mixture theory (CMT) of vascular G\&R. \cite{menzel2012frontiers} In CMT, families of structurally significant constituents are continuously deposited and degraded in response to the mechanobiological environment, while each deposited constituent is constrained to move with the mixture as a whole. CMT has been particularly successful at modeling vessel evolution in response to a wide variety of hemodynamic and pathological conditions. Recent developments have incorporated increasingly detailed biological processes, such as inflammation, into vascular G\&R models, enabling better modeling of vascular injuries and tissue-engineered vascular grafts. \cite{humphrey2021constrained,irons2020cell, humphrey2002constrained, szafron2018immuno,latorre2018modeling}

Because of its versatility and robustness, there has been persistent interest in using CMT to predict G\&R in patient-specific anatomies.\cite{figueroa2009computational, mousavi2019patient} 
 Computational fluid dynamics (CFD) tools have long existed that could readily supply the CMT with necessary hemodynamic values in arbitrarily complex geometries, typically defined within rigid wall domains. By contrast, fluid-solid interaction (FSI) frameworks couple fluid behavior with deformable solid wall geometries that exhibit specific material models. These tools have allowed for increasingly detailed investigations of hemodynamics in many clinical applications, including patient-specific simulations of coronary arteries, pulmonary arteries, and congenital heart lesions. \cite{schwarz2023beyond} When used in isolation, of course, both CFD and FSI simulations are limited to short term predictions. For this reason, there has been a call for melding FSI and G\&R models to yield so-called "fluid-solid-growth" (FSG) models capable of detailed long-term predictions. By enabling long-term predictions of changing hemodynamics and vascular wall G\&R, The incorporation of CMT into CFD simulations constitutes a critical step forward.  

The computational complexity of tracking the mass turnover and deformation history of the many cohorts of each constituent needed for the full CMT has prevented it from being coupled to full-fidelity FSI solvers. Instead, early implementations of CMT used reduced-order models of vessels to simplify computation. \cite{figueroa2009computational} More recently, multi-network and homogenized variations of the CMT have been introduced to increase computational tractability. However, neither consider the constituent-deposition history, and practical implementations have failed to utilize full fluid simulations for their mechano-mediated inputs. \cite{braeu2017homogenized, mousavi2019patient} An equilibrated formulation of CMT reduces computational cost by directly predicting long-term adaptations without considering intermediate evolution states, but while this is ideal for finding configurations that are at quasi steady-state, many clinically relevant problems hinge on finding configurations during periods of dynamic remodeling. \cite{latorre2018mechanobiologically, laubrie2022prestretch}  In addition, the equilibrated formulation, like the multi-network and homogenized formulations, has yet to be used with full-fidelity fluid simulations.  

\begin{figure}[!h]
\includegraphics[width=1.0\textwidth]{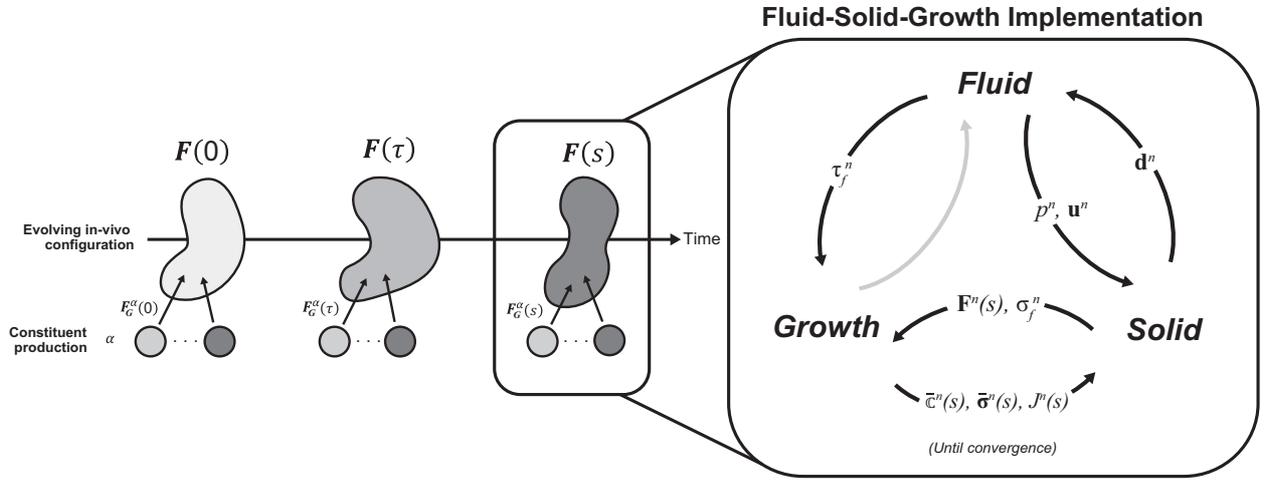}
\centering
\caption{Schematic view of the FSG framework where timestep updates are determined by iterative convergence of the fluid-structure-growth equations. The initial configuration of the mixture is chosen as the reference configuration for calculation of G\&R deformations of the mixture via $\mathbf{F}(\tau)$. The deformation experienced, at time $s$, by the material element of constituent $\alpha$ deposited at time $\tau$ is given by the deposition prestretch and subsequent deformation of the mixture expressed as $\mathbf{F}^\alpha_{n(\tau)}(s) = \mathbf{F}(s) \mathbf{F}^{-1}(\tau) \mathbf{F}^\alpha_G(\tau)$ where $\mathbf{F}(s) \mathbf{F}^{-1}(\tau)$ represents the constrained deformation due to the deformation of the mixture and $\mathbf{F}^\alpha_G(\tau)$ represents the deposition deformation gradient (which includes the deposition prestretch). At each timestep, we enforce convergence of the governing equations of the fluid, solid, and growth domains. To do this, we iteratively update the interfacing variables of each domain. This includes the fluid velocity, $\mathbf{u}^n$, wall shear stress, $\boldsymbol{\tau}^n_f$, and pressure, $p(s)$, from the fluid domain, the solid displacement, $\mathbf{d}^n$, deformation gradient, $\mathbf{F}^n(s)$, and the intramural Cauchy stress, $\boldsymbol{\sigma}^n_f$ from the solid domain, and the deformation-dependent spatial elasticity tensor, $\bar{\mathbb{c}}(s)$, and deformation-dependent Cauchy stress, $\bar{\boldsymbol{\sigma}}(s)$, from the growth domain. Note that there is not directly coupling from the growth domain to the fluid domain. Instead, convergence between the fluid and growth domains is indirectly mediated by the solid domain. The exact coupling algorithms and how these interfacing variables are utilized are described below.}
\label{fig:schamatic}
\end{figure}

To overcome the above challenges, our FSG framework couples the full CMT governing equations to a finite element solver that handles 3D FSI (Figure \ref{fig:schamatic}). We validate the performance of this framework against benchmark CMT problems and demonstrate its ability to handle both idealized and patient-specific geometries, including pathological hypertension and tissue-engineered vascular grafts. We describe implementation details of CMT unique to patient-specific formulations.

\section{Methods}

We begin by outlining the governing equations of the growth, fluid, and solid domains individually and then discuss the process for coupling them in a partitioned framework. We then provide implementation details and illustrative examples of specific uses of this framework.

\subsection{Constrained mixture formulation} \label{section:cmt}
Vascular tissues are comprised of myriad microscale constituents (collagen, elastin, glycosaminoglycans, smooth muscle cells, etc.) that display unique material behaviors. In CMT, any infinitesimal element of a reference volume, $\delta V_0$ inside the solid domain, $\Omega^s$, is composed of a spatially homogenized mixture of multiple constituents, $\alpha$, where the mass of each constituent at current time, $s$, is given as $\delta M^\alpha(s)$. From here, we follow the notation conventions set forth in Latorre et al. 2020.\cite{latorre2020fast} We note that while CMT has been previously used in low-dimensional idealized geometries, the following formulation is written fully 3D form.

The referential mass density of each constituent, $\rho^\alpha_R(s) = \delta M^\alpha(s)/\delta V_0$, can evolve over time and be expressed as
%
%
%
%
%
%
%
%
\begin{equation} \label{mass}
\rho^\alpha_R(s) =  \rho_R^\alpha(0)  Q^\alpha(s) + \int^s_0 m_R^\alpha(\tau)  q^\alpha(s,\tau) \dd\tau,
\end{equation}
where $\rho_R^\alpha(0)$ is the initial referential mass density of constituent $\alpha$, $Q^\alpha(s)$ is a mass survival function that represents the fraction of the initial mass cohort that survives to the current time $s$, $q^\alpha(s, \tau)$ is a mass survival function that represents the fraction of the mass deposited at past time $\tau$ that survives to current time $s$, and $m_R^\alpha(\tau)$ is the referential mass density production rate of constituent $\alpha$ given as
\begin{equation}
m^\alpha_R (\tau) = m^\alpha_N(\tau) \Upsilon^\alpha (\tau),
\end{equation}
where $m^\alpha_N (\tau)$ is a nominal production rate per unit reference volume that is mediated by a stimulus function $\Upsilon^\alpha (\tau)$. The referential strain energy, $W^\alpha_R = \delta W^\alpha(s)/\delta V_0$, can then be expressed as
\begin{equation}
W^\alpha_R (s) = \frac{\rho_R^\alpha(0) }{\hat{\rho}^\alpha} Q^\alpha(s) \hat{W}^\alpha(\mathbf{C}^\alpha_{n(0)} (s))  +  \int^s_{0} \frac{m^\alpha_R (\tau)}{\hat{\rho}^\alpha}   q^\alpha (s,\tau) \hat{W}^\alpha(\mathbf{C}^\alpha_{n(\tau)} (s)) \dd\tau,
\end{equation}
where $\hat{W}^\alpha(\mathbf{C}^\alpha_{n(\tau)} (s)) $ is a volume-specific strain energy function, $\hat{\rho}^\alpha$ is the intrinsic mass per unit volume of constituent $\alpha$, and $\mathbf{C}^\alpha_{n(\tau)} (s)$ is the right Cauchy-Green tensor obtained from the deformation gradient $\mathbf{F}^\alpha_{n(\tau)} (s)$ that represents the deformation experienced, at time $s$, by a material element of constituent $\alpha$ deposited at time $\tau$. $\mathbf{F}^\alpha_{n(\tau)} (s)$ is given as
\begin{equation}
\mathbf{F}^\alpha_{n(\tau)} (s) = \mathbf{F} (s)  \mathbf{F}^{-1}(\tau)  \mathbf{F}^\alpha_G(\tau),
\end{equation}
where $\mathbf{F}^\alpha_G(\tau)$ represents the constituent-specific deposition tensor at time $\tau$ and is defined as
\begin{equation}
\mathbf{F}^\alpha_G(\tau) = \mathbf{R} (\tau)  \left[\mathbf{R}^\text{T} (\tau)  \mathbf{G}^\alpha  \mathbf{R}(\tau) \right] = \mathbf{G}^\alpha  \mathbf{R}(\tau).
\end{equation}
Here, $\mathbf{G}^\alpha$ is a constant, symmetric, and volume-preserving deposition stretch tensor such that $\mathbf{G}^\alpha (\tau) = \mathbf{G}^\alpha = {\mathbf{G}^\alpha}^\text{T} = \textbf{constant} $ and $\text{det}(\mathbf{G}^\alpha) = 1$. Changes in deposition stretch angle that may arise as the vessel evolves are accounted for by $\mathbf{R}(\tau)$, the rotation tensor obtained from a polar decomposition of $\mathbf{F}(\tau)$. The remaining portion of $\mathbf{F}^\alpha_{n(\tau)} (s)$, expressed by $\mathbf{F} (s) \mathbf{F}^{-1}(\tau)$, accounts for the assumption that the motion of each constituent, once deposited, is constrained to that of the tissue as a whole, which is given by deformation gradient $\mathbf{F}(s)$. The full expression for the corresponding right Cauchy-Green deformation tensor is then
\begin{equation}\mathbf{C}^\alpha_{n(\tau)} (s) = {\mathbf{F}^\alpha_{n(\tau)}}^\text{T} (s)  \mathbf{F}^\alpha_{n(\tau)} (s)  = \left[ {\mathbf{F}^\alpha_G}^\text{T}(\tau) \mathbf{F}^{-\text{T}}(\tau) \right] \mathbf{C}(s) \left[ \mathbf{F}^{-1}(\tau)  \mathbf{F}^\alpha_G(\tau) \right],
\end{equation}
where $\mathbf{C}(s) = \mathbf{F}^\text{T} (s) \mathbf{F} (s)$ is a measurable, mixture level deformation. The referential strain energy of the mixture as whole is then given by a simple rule of mixtures as
\begin{equation}W_R (s) = \sum_\alpha W^\alpha_R (s).\end{equation}
The change in volume of the mixture, $J(s) = \delta V(s)/ \delta V_0$, can be expressed as
\begin{equation}J(s) = \text{det}(\mathbf{F}(s)) = \sum_\alpha  \rho^\alpha_R(s)/\hat{\rho}^\alpha, \end{equation}
where $J(s)$ satisfies both definitions set forth by the current deformation gradient and the evolving referential mass density.

The second Piola-Kirchhoff stress of constituent $\alpha$ is given as
\begin{equation} \mathbf{S}^\alpha(s) = 2 \frac{\partial W^\alpha_R(s)}{\partial \mathbf{C}(s)}.\end{equation}
It is convenient to use the chain rule to express the second Piola-Kirchhoff in terms of the constituent-specific right Cauchy-Green tensor,
\begin{equation} \mathbf{S}^\alpha(s) =  2 \left( \frac{\partial W^\alpha_R(s)}{\partial \mathbf{C}^{\alpha}_{n(\tau)}(s)} : \frac{\partial \mathbf{C}^{\alpha}_{n(\tau)}(s)}{\partial \mathbf{C}(s)} \right).\end{equation}
where the symbol $:$ is the double dot product operator. The double dot product of a second-order tensor with a fourth-order tensor is defined as $(\mathbf{A} : \mathbb{B})_{kl} = A_{ij}\mathbb{B}_{ijkl}$. The constituent-specific right Cauchy-Green tensor can be simplified to the form
\begin{equation}\mathbf{C}^{\alpha}_{n(\tau)}(s)  = \left[\left( {\mathbf{F}^\alpha_G}^\text{T}(\tau)   \mathbf{F}^{-\text{T}}(\tau) \right) \odot \left({\mathbf{F}^\alpha_G}^\text{T}(\tau)  \mathbf{F}^{-\text{T}}(\tau)\right) \right] : \mathbf{C}(s),\end{equation}
where the symbol $\odot$ is the mixed dyadic product operator $(\mathbf{A} \odot \mathbf{B})_{ijkl} = A_{ik}B_{jl}$. It follows that
\begin{equation}
\frac{\partial \mathbf{C}^{\alpha}_{n(\tau)}(s)}{\partial \mathbf{C}(s)} = \left( {\mathbf{F}^\alpha_G}^\text{T}(\tau)   \mathbf{F}^{-\text{T}}(\tau) \right) \odot  \left({\mathbf{F}^\alpha_G}^\text{T}(\tau) \mathbf{F}^{-\text{T}}(\tau) \right).
\end{equation}
We will further define a constituent-level second Piola-Kirchhoff stress as
\begin{equation}\hat{\mathbf{S}}^\alpha(\mathbf{C}^{\alpha}_{n(\tau)}(s)) = 2 \frac{\hat{W}^\alpha(\mathbf{C}^{\alpha}_{n(\tau)}(s))}{\mathbf{C}^{\alpha}_{n(\tau)}(s)}\end{equation}
so that
\begin{equation}
\mathbf{S}^\alpha(s) = \frac{1}{\hat{\rho}^\alpha} \left( \rho_R^\alpha(0) Q^\alpha(s)  \mathbf{G}^\alpha   \hat{\mathbf{S}}(\mathbf{C}^{\alpha}_{n(0)}(s))  \mathbf{G}^\alpha  + \int^s_0 m_R^\alpha(\tau) q^\alpha(s,\tau) \mathbf{F}^{-1}(\tau) \mathbf{F}^\alpha_G(\tau) \hat{\mathbf{S}}(\mathbf{C}^{\alpha}_{n(\tau)}(s)) {\mathbf{F}^\alpha_G}^\text{T}(\tau) \mathbf{F}^{-\text{T}}(\tau) \dd\tau \right)
\end{equation} 
The deformation-dependent (i.e. the part independent of reaction terms resulting from kinematic constraints) second Piola-Kirchhoff stress of the mixture as a whole is then
\begin{equation}
\bar{\mathbf{S}}(s) = \sum_\alpha \mathbf{S}^\alpha (s).
\end{equation}
The Cauchy stress of an individual constituent $\alpha$ can be obtained via a pushforward operator on $\mathbf{S}^\alpha(s)$,
\begin{equation}
\boldsymbol{\sigma}^\alpha (s) = \frac{1}{J(s)} \mathbf{F}(s) \mathbf{S}^\alpha (s) \mathbf{F}^\text{T}(s).
\end{equation}
This can also be written in terms of the constituent-level Cauchy stress as
\begin{equation}
\boldsymbol{\sigma}^\alpha (s) = \frac{1}{\hat{\rho}^\alpha}  \left( \rho^\alpha(0) Q^\alpha \hat{\boldsymbol{\sigma}}^\alpha(s,0) + \int^s_0 m^\alpha(\tau) q^\alpha(s,\tau) \hat{\boldsymbol{\sigma}}^\alpha(s,\tau) \dd\tau \right),
\end{equation}
where $m^\alpha (\tau) = m^\alpha_R(\tau)/J(\tau)$ is the mass production rate per unit of current volume of the mixture at time $\tau$ and
\begin{equation}
\hat{\boldsymbol{\sigma}}^\alpha(s,\tau) = \frac{1}{\text{det}(\mathbf{F}^\alpha_{n(\tau)}(s))} \mathbf{F}^\alpha_{n(\tau)}(s) \hat{\mathbf{S}}(\mathbf{C}^{\alpha}_{n(\tau)}(s)) \mathbf{F}^{\alpha T}_{n(\tau)}(s),
\end{equation}
where $\text{det}(\mathbf{F}^\alpha_{n(\tau)}(s)) = J(s)/J(\tau)$. The deformation-dependent Cauchy stress of the mixture is then
\begin{equation}
\bar{\boldsymbol{\sigma}}(s) = \sum_\alpha \boldsymbol{\sigma}^\alpha(s),
\end{equation}
and the full Cauchy stress for the mixture can be expressed as
\begin{equation}
\boldsymbol{\sigma}(s) = - p(s)\mathbf{I} + \bar{\boldsymbol{\sigma}}(s),
\end{equation}
where $p(s)$ is a pressure-type Lagrange multiplier that enforces the known volume ratio $J(s)$ from kinematics at time $s$. Further, because G\&R processes are considered to be slow compared to a cardiac cycle, evolving states can be computed in strong form via $\nabla \cdot \boldsymbol{\sigma} = \mathbf{0}$. For weak form solutions, however, it is useful to compute the stiffness matrices. The material elasticity tensor $\mathbb{C}^\alpha$ of constituent $\alpha$ that describes the change in stress which results from a change in strain is defined as
\begin{equation}
\mathbb{C}^\alpha = 2 \frac{\partial \mathbf{S}^\alpha (s)}{\partial \mathbf{C} (s)}.
\end{equation}
It is convenient to use the chain rule to express the material elasticity tensor in terms of the constituent-specific right Cauchy-Green tensor,
\begin{equation}
\mathbb{C}^\alpha(s) =  2 \left( \frac{\partial \mathbf{S}^\alpha (s)}{\partial \mathbf{C}^{\alpha}_{n(\tau)}(s)} : \frac{\partial \mathbf{C}^{\alpha}_{n(\tau)}(s)}{\partial \mathbf{C}(s)}  \right) =  4 \left( \frac{\partial \mathbf{C}^{\alpha}_{n(\tau)}(s)}{\partial \mathbf{C}(s)}: \frac{\partial^2 W^\alpha_R (s)}{\partial \mathbf{C}^{\alpha}_{n(\tau)}(s) \partial \mathbf{C}^{\alpha}_{n(\tau)}(s)} : \frac{\partial \mathbf{C}^{\alpha}_{n(\tau)}(s)}{\partial \mathbf{C}(s)}   \right).
\end{equation}
The double dot product of a fourth-order tensor with a fourth-order tensor is defined as $(\mathbb{A} : \mathbb{B})_{ijkl} = \mathbb{A} _{ijmn}\mathbb{B}_{mnkl}$. We can define a constituent-level material elasticity tensor as
\begin{equation}
\hat{\mathbb{C}}^\alpha(\mathbf{C}^{\alpha}_{n(\tau)}(s)) = 2 \frac{\hat{\mathbf{S}}^\alpha(\mathbf{C}^{\alpha}_{n(\tau)}(s))}{\mathbf{C}^{\alpha}_{n(\tau)}(s)} = 4  \frac{\partial^2 \hat{W}^\alpha (s)}{\partial \mathbf{C}^{\alpha}_{n(\tau)}(s) \partial \mathbf{C}^{\alpha}_{n(\tau)}(s)},
\end{equation}
so that
\begin{alignat}{2}
\mathbb{C}^\alpha(s) = &\frac{1}{\hat{\rho}^\alpha} \Bigg( \rho_R^\alpha(0) Q^\alpha  \left[ \mathbf{G}^\alpha \odot \mathbf{G}^\alpha \right] : \hat{\mathbb{C}}(\mathbf{C}^{\alpha}_{n(0)}(s)) : \left[ \mathbf{G}^\alpha : \mathbf{G}^\alpha \right]\\
& + \int^s_0 m_R^\alpha(\tau) q^\alpha(s,\tau) \bigg( \left[ \mathbf{F}^{-1}(\tau) \mathbf{F}^\alpha_G(\tau)\right] \odot \left[ \mathbf{F}^{-1}(\tau) 
 \mathbf{F}^\alpha_G(\tau)\right] \bigg) : \hat{\mathbb{C}}(\mathbf{C}^{\alpha}_{n(\tau)}(s)) : \bigg( \left[ {\mathbf{F}^\alpha_G}^\text{T}(\tau)  \mathbf{F}^{-\text{T}}(\tau)\right] \odot \left[ {\mathbf{F}^\alpha_G}^\text{T}(\tau)  \mathbf{F}^{-\text{T}}(\tau)\right] \bigg) \dd\tau \Bigg).   \nonumber
\end{alignat}
The spatial elasticity tensor of constituent $\alpha$, which corresponds to changes in Cauchy stress can then be obtained via a pushforward operator on $\mathbf{S}^\alpha(s)$,
\begin{equation}\mathbb{c}^\alpha(s) = \frac{1}{J(s)} \Big( \mathbf{F}(s) \odot \mathbf{F}(s) \Big) : \mathbb{C}^\alpha(s) : \Big( \mathbf{F}^\text{T}(s) \odot \mathbf{F}^\text{T}(s) \Big).\end{equation}
The deformation-dependent material elasticity tensor of the mixture is then 
\begin{equation}
\bar{\mathbb{C}}(s) = \sum_\alpha \mathbb{C}^\alpha(s),
\end{equation}
and the deformation-dependent spatial elasticity tensor of the mixture is 
\begin{equation}\bar{\mathbb{c}}(s) = \sum_\alpha \mathbb{c}^\alpha(s). \end{equation}
Specific choices for parameter values and functional forms are described in Section \ref{section:details}.

\subsection{Fluid formulation}
We define $\Omega^f$ as the fluid domain with $\Omega^f_0$ denoting the reference configuration of the fluid domain and $\Omega^f_\iota$ the current configuration.  For most studies of hemodynamics in large vessels, the governing equations for mass balance and linear momentum balance in a non-moving domain include the incompressible Navier-Stokes equations
\begin{alignat}{3} \label{ns}
\nabla \cdot \mathbf{u} &= 0 &\quad\text{ in } \Omega^f\\
\rho\dot{\mathbf{u}} + \nabla\cdot(\rho \mathbf{u} \otimes\mathbf{u} - \boldsymbol{\sigma}^f) - \rho \mathbf{f} &= \mathbf{0} &\quad\text{ in } \Omega^f,
\end{alignat}
where blood is assumed to exhibit a Newtonian behavior at high shear rates such that
\begin{equation}\boldsymbol{\sigma}^f = -p\mathbf{I} + \mu (\nabla \mathbf{u} + \nabla{\mathbf{u}}^\text{T})\end{equation}
is the fluid stress tensor, $\mathbf{u}$ is the fluid velocity, $\rho$ is the fluid density, $\mu$ is fluid viscosity, $p$ is pressure, $\mathbf{f}$ is a body force acting on the fluid, and $\nabla$ is the traditional differential operator acting on the current configuration unless otherwise noted. For moving domain simulations, it is necessary to pose these equations in a form amenable to discrete analysis in a moving domain. To do this, we can use a space-time Piola transformation to modify Equation~\eqref{ns} to the form
\begin{equation} \label{ns:3}
\rho \left( \dot{\mathbf{u}} + (\mathbf{u} - \hat{\mathbf{u}}) \cdot \nabla \mathbf{u} - \mathbf{f} \right) - \nabla \cdot \boldsymbol{\sigma}^f  = \mathbf{0},  \quad\text{ in } \Omega^f_\iota,
\end{equation}
where $\hat{\mathbf{u}}$ is the velocity of the fluid domain. Derivation details can be found in elsewhere. \cite{bazilevs2013computational} To solve the strong form of the fluid governing equations, we define the solution spaces $S_u$ and $S_p$ as well as the variational spaces $V_u$ and $V_p$.  We define the boundary of the fluid domain as $\Gamma_f$ and assert the existence of $\Gamma^f_h$ and $\Gamma^f_g$ which refer to the portion of the boundary with natural and essential boundary conditions respectively. The boundary conditions are then defined on the solution space as
\begin{alignat}{3}
u &= g \text{ on } \Gamma^f_g \text{, }& \quad\forall u \in S\\
w &= 0 \text{ on } \Gamma^f_g \text{, }& \quad\forall w \in V , 
\end{alignat}
where $u$ and $w$ are members of the function sets $S$ and $V$ respectively, and $g$ is a given function.

We then pose the variational form of the incompressible, moving-domain Navier Stokes equations. Find $\mathbf{u} \in S_u$ and $p \in S_p$ such that for all $\mathbf{w} \in V_u$ and $q \in V_p$ the following is satisfied:
\begin{equation}
\int_{\Omega^f_\iota} \mathbf{w} \cdot \rho \left( \dot{\mathbf{u}} + (\mathbf{u} - \hat{\mathbf{u}}) \cdot \nabla \mathbf{u} - \mathbf{f} \right) \dd\Omega + \int_{\Omega^f_\iota} \epsilon(\mathbf{w}) : \sigma^f (\mathbf{u},p) \dd\Omega - \int_{\Gamma^f_{h_\iota}} \mathbf{w} \cdot \mathbf{h} \dd \Gamma_h - \int_{\Omega^f_\iota} q \nabla \cdot \mathbf{u} \dd\Omega = \mathbf{0}.
\end{equation}
Additional implementation details of the discretization and stabilization of the fluid formulation can be found in \ref{app:fluid}. In the end, we are left with an iterative matrix system for fluid mechanics given in terms of $\dot{\mathbf{u}}_{n+1}$ and $p_{n+1}$ that can be expressed in the form

\begin{equation}
\left( \begin{matrix} 
\mathbf{K} & \mathbf{G} \\
\mathbf{D} & \mathbf{L} \\
\end{matrix} \right)
\left( \begin{matrix} 
\Delta \dot{\mathbf{u}}_{n+1} \\
\Delta p_{n+1} \\
\end{matrix} \right)
=
\left( \begin{matrix} 
-\mathbf{R}^{mom} \\
-\mathbf{R}^{cont} \\
\end{matrix}\right),
\end{equation}

where $\mathbf{R}^{cont}$ and $\mathbf{R}^{mom}$ are the residuals of the continuity and linear momentum equations, respectively, and $\mathbf{K}$, $\mathbf{G}$, $\mathbf{D}$, and $\mathbf{L}$ are given as
\begin{alignat}{2}
\mathbf{K} &= \frac{\partial \mathbf{R}^{mom}}{\partial \dot{\mathbf{u}}_{n+1}}\\
\mathbf{G} &= \frac{\partial \mathbf{R}^{mom}}{\partial p_{n+1}}\\
\mathbf{D} &= \frac{\partial \mathbf{R}^{cont}}{\partial \dot{\mathbf{u}}_{n+1}}\\
\mathbf{L} &= \frac{\partial \mathbf{R}^{cont}}{\partial p_{n+1}}.
\end{alignat}

\subsection{Solid formulation}

 In $\Omega^s$ the governing equations must similarly satisfy linear momentum balance,
\begin{equation} \label{me}
\mathrm{div } \boldsymbol{\sigma} + \rho (\mathbf{f} - \mathbf{a}) = \mathbf{0} \quad\text{ in } \Omega^s_\iota,
\end{equation}
where $\boldsymbol{\sigma}$ is the Cauchy stress in the solid, $\mathbf{f}$ is the solid body force per unit mass, $\rho$ is the solid mass density, and $\mathbf{a}$ is the solid acceleration. 

Structural mechanics are solved computationally using similar methods. Here, we denote a reference and current domain, denoted $\Omega_0$ and $\Omega_\iota$, respectively. We define our trial and test function spaces as
\begin{alignat}{3}
S_\mathbf{d} &= \{\mathbf{d} | \mathbf{d}(\cdot, \iota) \in (H^1(\Omega_\iota))^{n_{sd}}, d_i = g_i &\quad\text{ on } (\Gamma^s_\iota)_{g_i}\}\\
V_\mathbf{d} &= \{\mathbf{w} | \mathbf{w}(\cdot, \iota) \in (H^1(\Omega_\iota))^{n_{sd}}, w_i = 0 &\quad\text{ on } (\Gamma^s_\iota)_{g_i}\}.
\end{alignat}
The weak formulation of the linear momentum equation is then
\begin{equation}
\int_{\Omega^s_\iota} \mathbf{w} \rho \cdot \mathbf{f} \dd \Omega - \int_{\Omega^s_\iota} \mathbf{w} \rho \cdot \ddot{\mathbf{d}} \dd\Omega + \int_{\Gamma^s_{\iota_h}} \mathbf{w} \cdot \mathbf{h} \dd \Gamma - \int_{\Omega^s_0} \delta \mathbf{S}:\mathbf{E} \dd\Omega = \mathbf{0},
\end{equation}
which can be expressed solely in the current configuration as
\begin{equation}
\int_{\Omega^s_\iota} \mathbf{w} \rho \cdot \mathbf{f} \dd \Omega - \int_{\Omega^s_\iota} \mathbf{w} \rho \cdot \ddot{\mathbf{d}} \dd\Omega + \int_{\Gamma^s_{\iota_h}} \mathbf{w} \cdot \mathbf{h} \dd \Gamma - \int_{\Omega^s_\iota} \boldsymbol{\sigma}:\epsilon(\mathbf{w}) \dd\Omega = \mathbf{0}.
\end{equation}
Additional implementation details of the discretization of the solid formulation can be found in \ref{app:solid}. In the end, we are left with an iterative matrix system for structural mechanics given in terms of $\ddot{\mathbf{d}}_{n+1}$,

\begin{equation}
\mathbf{K}\Delta\ddot{\mathbf{d}}_{n+1,i} = -\mathbf{R}^{str}_{i-1}.
\end{equation}

where $\mathbf{R}^{str}$ is the residual of the linear momentum equation and $\mathbf{K}$ is given as

\begin{equation}
    \mathbf{K} = \frac{\partial \mathbf{R}^{str}}{\partial \ddot{\mathbf{d}}_{n+1}}
\end{equation}

This fully defines our solid formulation and allows us to solve for displacement and, by extension, the deformation gradient and Cauchy stress in the solid domain.

\subsection{Fluid-solid-growth coupling}

When coupling the equations of the fluid domain ($\mathbf{\mathcal{F}}$), the solid domain ($\mathbf{\mathcal{S}}$), and the growth domain ($\mathbf{\mathcal{G}}$), it is crucial to consider the different timescales involved in each domain. Depending on the specific biological processes being modeled, the changes over $\mathbf{\mathcal{G}}$ occur over periods ranging from days to months or even years. By contrast, the response times of the hyperelastic material in the solid domain and the viscous material in the fluid domain are much shorter, typically on the scale of milliseconds or seconds. Therefore, we make the assumption that the growth domain changes at a much slower rate compared to the fluid and solid domains such that
\begin{equation}\mathbf{\mathcal{G}} = \mathbf{constant} \quad\text{ on } \Omega^s \text{ and } \Omega^f\end{equation}
for the duration of time-stepped simulations in the fluid and solid domains. Under this assumption, we can choose a timescale to use with our fluid and solid simulations that is independent from the timescale of our G\&R domain. We can then evaluate the growth domain at discrete timesteps such that time $s$ can be expressed as
\begin{equation}s = \Delta s + \tau,\end{equation}
where $\tau$ is a previous time and $\Delta s$ is a discrete timestep. At each of these discrete timesteps, the inputs from the fluid and solid domains into the growth domain can be calculated and the integrals in the growth domain can then be evaluated using a numerical integration scheme. For example, using trapezoidal integration,

\begin{equation}
\rho^\alpha_R(s) = \Delta s \frac{1}{2}\rho_R^\alpha(0)  Q^\alpha(s) +  \left(
 \Delta s \sum^{n_{\Delta_s} - 1}_{k=1} m_R^\alpha(s - k\Delta s)  q^\alpha(s,s - k\Delta s) \right) + \Delta s \frac{1}{2} m_R^\alpha(s) q^\alpha(s,s),
\end{equation}

is the discretized form of the mass density function where $n_{\Delta_s}$ is the number of timesteps at a simulated time of $s = n_{\Delta_s}\Delta s$. The other integrals in the growth domain can be similarly discretized.

To simultaneously find the configuration that satisfies the governing equations of the growth domain, $\mathbf{\mathcal{G}}$, the fluid domain, $\mathbf{\mathcal{F}}$, and the solid domain, $\mathbf{\mathcal{S}}$ at each timestep, we propose a partitioned approach that iteratively couples the fluid-solid and solid-growth domains until convergence of all respective formulations. The details of the fluid-solid and solid-growth coupling are given below.

The most general form of FSG iterative coupling is given in Algorithm \ref{pulsatile_coupling}. Its use of use of monolithic fluid-solid coupling at every FSG iteration is generic to both pulsatile and steady-state flow simulations. We also introduce two variations of this coupling schemes described in Algorithm \ref{fsg_coupling} and \ref{alt_fsg_coupling}. While all algorithms  iterate between the fluid-solid and solid-growth coupling and enforce convergence of the respective governing equations, Algorithm \ref{fsg_coupling} utilizes partitioned fluid-solid coupling and is optimized for problems where the fluid is in steady-state. Algorithm \ref{alt_fsg_coupling} also utilizes partitioned fluid-solid coupling for steady-state fluid problems, but enforces convergence of the solid-growth coupling before iterating the fluid-solid coupling. This minimizes the number of times the fluid solver is called and is more efficient in problems where the expected evaluation time of the fluid framework is much larger than that of the solid framework or growth framework, such as in complex geometries with large fluid domains.




\subsubsection{Fluid-solid coupling}

We define the interface of the fluid and solid domains as
\begin{equation}
\Gamma^I = \Gamma^f \bigcap \Gamma^s,
\end{equation}
where $\Gamma^f$ is the fluid boundary and $\Gamma^s$ is the structural boundary. To ensure that the fluid and structure domains remain physically interfaced, we enforce the interface boundary conditions
\begin{alignat}{3}
\mathbf{d}^f &= \mathbf{d}^s &\quad\text{ on } \Gamma^I\\
\mathbf{u}^f &= \dot{\mathbf{d}}^s &\quad\text{ on } \Gamma^I\\
\mathbf{w}^f &= \mathbf{w}^s &\quad\text{ on } \Gamma^I\\
\boldsymbol{\sigma}^f \hat{\mathbf{n}}^f &= -\boldsymbol{\sigma}^s \hat{\mathbf{n}}^s  &\quad\text{ on } \Gamma^I,
\end{alignat}
where $\mathbf{w}^f$ and $\mathbf{w}^s$ are the test functions of the fluid and structural domains, $\boldsymbol{\sigma}^f$ and $\boldsymbol{\sigma}^s$ are the current Cauchy stresses of the fluid and structural domains, and $\hat{\mathbf{n}}^f$ and $\hat{\mathbf{n}}^s$ are the normal vectors of the fluid and structural domains.

In this study, we use a partitioned scheme to solve steady-state FSI problems and a monolithic scheme to solve unsteady or pulsatile FSI problems. This allows flexibility in building our overall FSG coupling scheme.

In steady-state problems,
\begin{equation}\mathbf{u}^f = \dot{\mathbf{d}}^s = \mathbf{0} \quad\text{ on } \Gamma^I.\end{equation}
To enforce this, we generate a volumetric fluid mesh directly from the enforced $\Gamma^I$ and enforce $\mathbf{u}^f = \mathbf{0}$ on $\Gamma^I$. We then run a rigid-wall fluid simulation to yield the pressure field and wall shear stress on $\Gamma^I$:
\begin{equation}
p^n, \mathbf{u}^n = \mathbf{\mathcal{F}}(\mathbf{d^n}),
\end{equation}
where the superscript $n$ respresents the $n$th coupling iteration between the fluid and solid domains. Given this solution, we obtain an updated displacement field by running a solid simulation using the Gauss-Seidel scheme such that
\begin{equation}
\mathbf{d}^{n+1} = \mathbf{\mathcal{S}}(p^n, \mathbf{u}^n).
\end{equation}

In pulsatile simulations, we simultaneously solve for the fluid and solid domains using monolithic coupling by assembling the matrix
\begin{equation}
\left( \begin{matrix} 
\mathbf{K}_{11} & \mathbf{G}_{1} & \mathbf{K}_{12} & \mathbf{0}\\
\mathbf{D}_1 & \mathbf{L} & \mathbf{D}_2 & \mathbf{0}\\
\mathbf{K}_{21} & \mathbf{G}_2 & \mathbf{K}_{22} & \mathbf{0}\\
\mathbf{0} & \mathbf{0} & \mathbf{0} & \mathbf{K}_{33}\\
\end{matrix} \right)
\left( \begin{matrix} 
\Delta \dot{\mathbf{u}}_{n+1} \\
\Delta p_{n+1} \\
\Delta \ddot{\mathbf{d}}_{n+1} \\
\Delta \dot{\hat{\mathbf{u}}}_{n+1}  \\
\end{matrix} \right)
=
\left( \begin{matrix} 
-\mathbf{R}^{mom} \\
-\mathbf{R}^{cont} \\
-\mathbf{R}^{str} \\
-\mathbf{R}^{msh} \\
\end{matrix}\right),
\end{equation}

where $\mathbf{K}_{12}$, $\mathbf{D}_2$, $ \mathbf{K}_{21}$, and $\mathbf{G}_2$ can be calculated from the interface boundary conditions, and $\mathbf{R}^{msh}$ and $\mathbf{K}_{33}$ represent the system of equations governing the linear elastic movement of the fluid domain mesh. Then,

\begin{equation}
\mathbf{d}^{n+1},p^{n+1}, \mathbf{u}^{n+1} = \mathbf{\mathcal{F}}\mathbf{\mathcal{S}}(\mathbf{d}^n, p^n, \mathbf{u}^n),
\end{equation}

where $\mathbf{\mathcal{F}}\mathbf{\mathcal{S}}$ is the monolithically coupled fluid-solid domain.

\subsubsection{Solid-growth coupling}

A naive choice of partitioning method would be to couple the solid and growth equations such that
\begin{equation}
\boldsymbol{\sigma}^{n+1}(s) = \mathbf{\mathcal{G}}(\mathbf{d}^n)
\end{equation}
\begin{equation}
\mathbf{d}^{n+1} = \mathbf{\mathcal{S}}(\boldsymbol{\sigma}^{n+1}(s)).
\end{equation}
We note that the solution field from the fluid domain, $\mathbf{u}^n$, is an input into $\mathbf{\mathcal{G}}$ as the calculated hemodynamics in $\mathbf{\mathcal{F}}$ mediate G\&R via the mass production and survival functions. However, the growth domain's influence on the fluid domain is mediated via the solid domain as there are no inputs into the fluid equations that directly stem from the growth domain solution. Therefore, we focus on coupling the fluid-solid and solid-growth domains separately.

Here, we propose an alternate form of $\boldsymbol{\sigma}^{n+1}(s)$ to achieve a solid-growth coupling framework that solely utilizes the deformation-dependent material stiffness tensor and allows efficient implementation inside the solid formulation framework. To evolve a vessel configuration from an intermediate time, $\tau$, to a current time, $s$, we initialize the iterative process by setting
\begin{equation}
\postscript{n}{\mathbf{F}}(s) = \mathbf{F}(\tau),
\end{equation}
where the superscript $n$ represents the $n$th iteration at time $s$. Under this assumption, we then calculate $\postscript{n}{\Upsilon^{\alpha,}}(s)$, $\postscript{n}{\hat{\boldsymbol{\sigma}}}(s)$, $\postscript{n}{\mathbb{c}}(s)$. Given $\postscript{n}{\Upsilon^{\alpha,}}(s)$, we can calculate the expected volume update from the mass density function as 
\begin{equation}
\postscript{n+1}{J}(s) = \sum_\alpha \postscript{n}{\rho^{\alpha,}_R}(s) / \hat{\rho}^\alpha,
\end{equation}
which will not necessarily equal the volume at the current iteration given as
\begin{equation}
\postscript{n}{J}(s) = \text{det}(\postscript{n}{\mathbf{F}}(s)),
\end{equation}
but will converge as part of the iterative process. If $\postscript{n}{\mathbf{F}}(s)$ is expected to be a reasonable estimate of $\postscript{n+1}{\mathbf{F}}(s)$, the change in configuration under the solid formulation, 
$\postscript{*}{\mathbf{F}} = \postscript{n+1}{\mathbf{F}}(s) \postscript{n}{\mathbf{F}}(s)^{-1}$, can be expected to be reasonably small. Under this assumption, the behavior of the material under $\postscript{*}{\mathbf{F}}$ can be well approximated in its linearized regime. Note that the assumption of a "reasonably" small deformation gradient is distinct from the assumption of an "infinitesimally" small deformation gradient where higher order terms are ignored. The Cauchy stress at the $n+1$ iteration is expressed as 
\begin{equation}
\postscript{n+1}{\boldsymbol{\sigma}}(s) = -\postscript{n+1}{p}(s)  \mathbf{I} + \postscript{n+1}{\bar{\boldsymbol{\sigma}}}(s).
\end{equation}
Assuming the stress response in $\postscript{*}{\mathbf{F}}$ is nearly linear, we approximate the evolution of $ \postscript{n+1}{\boldsymbol{\sigma}}(s)$ as $\postscript{n+1}{\boldsymbol{\sigma}}(s) = \postscript{n}{\boldsymbol{\sigma}}(s) + \postscript{*}{\boldsymbol{\sigma}}$ and rewrite $\postscript{n+1}{\boldsymbol{\sigma}}(s)$ as
\begin{equation}
\postscript{n+1}{\boldsymbol{\sigma}}(s) = -(\postscript{n}{p}(s) + \postscript{*}{p})\mathbf{I} + \frac{1}{\postscript{n}{J}(s) \postscript{*}{J} } \postscript{*}{\mathbf{F}} \postscript{n}{\mathbf{F}}(s)(\postscript{n}{\bar{\mathbf{S}}}(s) + \postscript{*}{\bar{\mathbf{S}}}) \postscript{n}{{\mathbf{F}}}(s)^\text{T}{\postscript{*}{\mathbf{F}}}^\text{T},
\end{equation}
where $\postscript{*}{J} = \postscript{n+1}{J}(s)/\postscript{n}{J}(s)$. We approximate $\postscript{*}{\bar{\mathbf{S}}}$ by linearizing the stress response at the current iteration
\begin{equation}
\postscript{*}{\bar{\mathbf{S}}} = \postscript{n}{\bar{\mathbb{C}}}(s) \postscript{*}{\mathbf{H}},
\end{equation}
where
\begin{equation}
\postscript{*}{\mathbf{H}} = {\postscript{n}{\mathbf{F}}}(s)^\text{T} ({\postscript{*}{\mathbf{C}}}- \mathbf{I}) \postscript{n}{\mathbf{F}}(s).
\end{equation}
$\postscript{n}{\bar{\boldsymbol{\sigma}}}(s)$ can be substituted in for $\postscript{n}{\mathbf{F}}(s)\postscript{n}{\mathbf{S}}(s)\postscript{n}{{\mathbf{F}}}(s)^\text{T}$ and the equations simplifies to
\begin{equation}
\postscript{n+1}{\boldsymbol{\sigma}}(s) = -(\postscript{n}{p}(s) + \postscript{*}{p})\mathbf{I} + \frac{1}{J^*} \postscript{*}{\mathbf{F}}\postscript{n}{\bar{\boldsymbol{\sigma}}}(s){\postscript{*}{\mathbf{F}}}^\text{T} + \frac{1}{J^* J^n(s)}\postscript{*}{\mathbf{F}} \postscript{n}{\mathbf{F}}(s)(\postscript{n}{\bar{\mathbb{C}}}(s) \postscript{*}{\mathbf{H}})\postscript{n}{\mathbf{F}}(s)^\text{T}{\postscript{*}{\mathbf{F}}}^\text{T}.
\end{equation}
Defining the Green-Lagrange strain tensor, $\postscript{*}{\mathbf{E}} = \frac{1}{2}\left( \postscript{*}{\mathbf{C}}- \mathbf{I} \right) $, and transitioning to use the spatial material elasticity tensor, we produce the full form of $\postscript{n+1}{\boldsymbol{\sigma}}(s)$ as
\begin{equation}
\postscript{n+1}{\boldsymbol{\sigma}}(s) = -(\postscript{n}{p}(s) + \postscript{*}{p})\mathbf{I} + \frac{1}{\postscript{*}{J}} \postscript{*}{\mathbf{F}} \postscript{n}{\bar{\boldsymbol{\sigma}}}(s){\postscript{*}{\mathbf{F}}}^\text{T} + \frac{1}{\postscript{*}{J}}\postscript{*}{\mathbf{F}} (\postscript{n}{\bar{\mathbb{c}}}(s):\postscript{*}{\mathbf{E}}){\postscript{*}{\mathbf{F}}}^\text{T}.
\end{equation}
We use a dilational penalty framework to enforce the target volume $\postscript{n+1}{J}(s)$ and define $\postscript{*}{p}(s)$ such that
\begin{equation}
\postscript{*}{\mathbf{E}} =\frac{1}{2}\left[\left(\frac{\text{det}( \postscript{*}{\mathbf{F}}) }{\postscript{*}{J}}\right)^{-2/3}{\postscript{*}{\mathbf{F}}}^\text{T} \postscript{*}{\mathbf{F}} - \mathbf{I} \right]
\end{equation}
\begin{equation}
\postscript{*}{p} = -k_p(\text{det} (\postscript{*}{\mathbf{F}} )- \postscript{*}{J}),
\end{equation}
where $k_p$ is a volumetric penalty parameter.
This stress equation has the analogous form to the St. Venant-Kirchhoff material with prestress and a penalty enforcement which is easily implemented in many standard solid mechanics solvers.
The final configuration is then $\postscript{n+1}{\mathbf{F}}$ with the new displacement field $\mathbf{d}^{n+1}$. To aid in the convergence of the solid-growth coupling, we additionally utilize Aitken relaxation where
\begin{equation}
\tilde{\mathbf{d}}^{n+1} = \mathbf{\mathcal{S}} \circ \mathbf{\mathcal{G}}(\mathbf{d}^n),
\end{equation}
giving $\tilde{\mathbf{d}}^{n+1}$ as our unrelaxed displacement and
\begin{equation}
\mathbf{r}^{n} = \tilde{\mathbf{d}}^{n+1}-\mathbf{d}^{n}
\end{equation}
as the displacement residual. The updated displacement can then be expressed as
\begin{equation}
\mathbf{d}^{n+1} = \omega^{n}\tilde{\mathbf{d}}^{n+1} + (1 - \omega^{n})\mathbf{d}^{n},
\end{equation}
where
\begin{equation}
\omega^{n} = - \omega^{n-1}\frac{(\mathbf{r}^{n-1})^\text{T}(\mathbf{r}^{n} - \mathbf{r}^{n-1})}{(\mathbf{r}^{n} - \mathbf{r}^{n-1})^\text{T} (\mathbf{r}^{n} - \mathbf{r}^{n-1})}.
\end{equation}

With the solid-growth coupling and fluid-solid coupling defined, we iterate between these coupling schemes until convergence of the three domains.

\begin{algorithm}
\begin{algorithmic}[1]
\State $t=0, ~k=0$
\State Initialize geometry utilizing full mixture model
\While{$t \leq t_\text{max}$}
    \While{$n \leq n_\text{max}$}
        \State $\tilde{\mathbf{d}}^{n+1} = \mathbf{\mathcal{F}}{\mathcal{S}}(\mathbf{d}^n, \boldsymbol{\sigma}^{n}(s))$
        \If{$n \leq 1$}
            \State $\omega^n = 0.5$
        \Else
            \State $\omega^{n} = - \omega^{n-1}\frac{(\mathbf{r}^{n-1})^\text{T}(\mathbf{r}^{n} - \mathbf{r}^{n-1})}{(\mathbf{r}^{n} - \mathbf{r}^{n-1})^\text{T} (\mathbf{r}^{n} - \mathbf{r}^{n-1})}$
        \EndIf
        \State $\mathbf{d}^{n+1} = \omega^{n}\tilde{\mathbf{d}}^{n+1} + (1 - \omega^{n})\mathbf{d}^{n}$
        \State $\boldsymbol{\sigma}^{n+1}(s) = \mathbf{\mathcal{G}}(\mathbf{d}^{n+1})$

        \State $r_{tol} = \Vert \mathbf{r} \Vert_\infty$
        \State $\tau_{tol} = \Vert | \tau^n - \tau^{n-1} | / \tau_h\Vert_\infty$
        \State $\sigma_{tol} = \Vert | \sigma^n - \sigma^{n-1} | / \sigma_h\Vert_\infty$
        \State $J_{tol} = \Vert | J^n/J^{n-1} - 1 |\Vert_\infty$
        
        \If{$r_{tol} < \epsilon_r$ and $\tau_{tol} < \epsilon_\tau$ and $\sigma_{tol} < \epsilon_\sigma$ and $J_{tol} < \epsilon_J$ }
            \State break
        \EndIf
        \State $n++$
    \EndWhile
    \State $t++$
    \State $n = 0$
\EndWhile
\end{algorithmic}
\caption{The general form of a fluid-solid-growth coupling scheme that utilizes monolithic fluid-solid coupling and is generic to both pulsatile and steady-state flow simulations. \label{pulsatile_coupling}}
\end{algorithm}
\begin{algorithm}
\begin{algorithmic}[1]
\State $t=0, ~n=0$
\State Initialize geometry utilizing full mixture model
\While{$t \leq t_\text{max}$}
    \While{$n \leq n_\text{max}$}
        \State $p^n = \mathbf{\mathcal{F}}(\mathbf{d}^{n})$
        \State $\tilde{\mathbf{d}}^{n+1} = \mathbf{\mathcal{S}}(p^n, \boldsymbol{\sigma}^{n}(s))$
        \If{$n \leq 1$}
            \State $\omega^n = 0.5$
        \Else
            \State $\omega^{n} = - \omega^{n-1}\frac{(\mathbf{r}^{n-1})^\text{T}(\mathbf{r}^{n} - \mathbf{r}^{n-1})}{(\mathbf{r}^{n} - \mathbf{r}^{n-1})^\text{T} (\mathbf{r}^{n} - \mathbf{r}^{n-1})}$
        \EndIf
        \State $\mathbf{d}^{n+1} = \omega^{n}\tilde{\mathbf{d}}^{n+1} + (1 - \omega^{n})\mathbf{d}^{n}$
        \State $\boldsymbol{\sigma}^{n+1}(s) = \mathbf{\mathcal{G}}(\mathbf{d}^{n+1})$

        \State $r_{tol} = \Vert \mathbf{r} \Vert_\infty$
        \State $\tau_{tol} = \Vert | \tau^{n+1}_f - \tau^n_f | / \tau_h\Vert_\infty$
        \State $\sigma_{tol} = \Vert | \sigma^{n+1}_f - \sigma^{n}_f | / \sigma_h\Vert_\infty$
        \State $J_{tol} = \Vert | J^{n+1}/J^{n} - 1 |\Vert_\infty$
        
        \If{$r_{tol} < \epsilon_r$ and $\tau_{tol} < \epsilon_\tau$ and $\sigma_{tol} < \epsilon_\sigma$ and $J_{tol} < \epsilon_J$ }
            \State break
        \EndIf
        \State $n++$
    \EndWhile
    \State $t++$
    \State $n=0$
\EndWhile
\end{algorithmic}
\caption{Fluid-solid-growth coupling scheme for steady flow problems that utilize partitioned fluid-solid coupling. \label{fsg_coupling}}
\end{algorithm}
\begin{algorithm}
\begin{algorithmic}[1]
\State $t=0, ~k=0$
\State Initialize geometry utilizing full mixture model
\While{$t \leq t_\text{max}$}
    \While{$k \leq k_\text{max}$}
        \If{k = 0}
            \State $p^n = \mathbf{\mathcal{F}}(\mathbf{d}^{n})$
        \EndIf
        \While{$j \leq j_\text{max}$}
            \State $\tilde{\mathbf{d}}^{n+1} = \mathbf{\mathcal{S}}(p^n, \boldsymbol{\sigma}^{n}(s))$
            \If{$n \leq 1$}
                \State $\omega^n = 0.5$
            \Else
                \State $\omega^{n} = - \omega^{n-1}\frac{(\mathbf{r}^{n-1})^\text{T}(\mathbf{r}^{n} - \mathbf{r}^{n-1})}{(\mathbf{r}^{n} - \mathbf{r}^{n-1})^\text{T} (\mathbf{r}^{n} - \mathbf{r}^{n-1})}$
            \EndIf
            \State $\mathbf{d}^{n+1} = \omega^{n}\tilde{\mathbf{d}}^{n+1} + (1 - \omega^{n})\mathbf{d}^{n}$
            \State $\boldsymbol{\sigma}^{n+1}(s) = \mathbf{\mathcal{G}}(\mathbf{d}^{n+1})$
    
            \State $r_{tol} = \Vert \mathbf{r} \Vert_\infty$
            \State $\sigma_{tol} = \Vert | \sigma^{n+1}_f - \sigma^{n}_f | / \sigma_h\Vert_\infty$
            \State $J_{tol} = \Vert | J^{n+1}/J^{n} - 1 |\Vert_\infty$
            \State n++
            \If{$r_{tol} < \epsilon_r$ and $\sigma_{tol} < \epsilon_\sigma$ and $J_{tol} < \epsilon_J$ }
                \State break
            \EndIf
        \EndWhile
        \State $p^n = \mathbf{\mathcal{F}}(\mathbf{d}^{n})$
        \State $\tilde{\mathbf{d}}^{n+1} = \mathbf{\mathcal{S}}(p^n, \boldsymbol{\sigma}^{n}(s))$
        \If{$n \leq 1$}
            \State $\omega^n = 0.5$
        \Else
            \State $\omega^{n} = - \omega^{n-1}\frac{(\mathbf{r}^{n-1})^\text{T}(\mathbf{r}^{n} - \mathbf{r}^{n-1})}{(\mathbf{r}^{n} - \mathbf{r}^{n-1})^\text{T} (\mathbf{r}^{n} - \mathbf{r}^{n-1})}$
        \EndIf
        \State $\mathbf{d}^{n+1} = \omega^{n}\tilde{\mathbf{d}}^{n+1} + (1 - \omega^{n})\mathbf{d}^{n}$
        \State $\boldsymbol{\sigma}^{n+1}(s) = \mathbf{\mathcal{G}}(\mathbf{d}^{n+1})$

        \State $r_{tol} = \Vert \mathbf{r} \Vert_\infty$
        \State $\tau_{tol} = \Vert | \tau^{n+1}_f - \tau^n_f | / \tau_h\Vert_\infty$
        \State $\sigma_{tol} = \Vert | \sigma^{n+1}_f - \sigma^{n}_f | / \sigma_h\Vert_\infty$
        \State $J_{tol} = \Vert | J^{n+1}/J^{n} - 1 |\Vert_\infty$
        
        \If{$r_{tol} < \epsilon_r$ and $\tau_{tol} < \epsilon_\tau$ and $\sigma_{tol} < \epsilon_\sigma$ and $J_{tol} < \epsilon_J$ }
            \State break
        \EndIf
        \State $k++$
        \State $n++$
        \State $j = 0$
    \EndWhile
    \State $t++$
    \State $n=0$
\EndWhile
\end{algorithmic}
\caption{Fluid-solid-growth coupling scheme for steady flow problems that utilize with partitioned fluid-solid coupling. Additionally, the solid-growth coupling is enforced to converge prior to iterating the fluid-solid coupling. \label{alt_fsg_coupling}}
\end{algorithm}

\section{Implementation Details}
\label{section:details}

The finite-element formulation of constrained mixture G\&R was implemented in the open-source solver svFSI, which is part of the SimVascular software suite.\cite{updegrove16} Distance-based linear hexehedral C3D8 elements were used for spatial discretization unless otherwise noted.\cite{lapidus2011numerical} Deformation gradient histories were tracked at every Gauss point while hemodynamic inputs into the $\Upsilon(s)$ functions were spatially smoothed using a distance-weighted Laplacian kernel and cell-wise averaged. The trapezoidal rule was used for numerical integration of the heredity integrals in the growth domain at discrete timesteps. At $s=0$ the full constrained mixture model was used to initialize the model geometry. At the beginning of each solid iteration, a ramping function and mass dampening factor was used to smooth the transition from $\mathbf{F}^n(s)$ to $\mathbf{F}^{n+1}(s)$. 

\subsection{Constituent strain energy}

In our implementation, we model biological constituents as either isotropic or anisotropic. For isotropic constituents (elastin, ground matrix, and polymer), the stored energy is calculated using a neoHookean model:
\begin{equation}
\hat{W}^{\alpha}(\mathbf{C}^{\alpha}_{n(\tau)}(s)) = \frac{c^{\alpha}}{2}(\mathbf{C}^{\alpha}_{n(\tau)}(s):\mathbf{I}-3).
\end{equation}
Anisotropic constituents (collagen fibers, passive smooth muscle) are assumed to have an associated preferential fiber direction denoted by unit vector $\mathbf{h}^\alpha(\tau) = \mathbf{R}(\tau) \mathbf{h}^\alpha_0$ where $\mathbf{h}^\alpha_0$ is the fiber direction of constituent $\alpha$ in the reference configuration. The associated stretch in the direction of this fiber direction is then given as the pseudo-invariant $\sqrt{\mathbf{C}^{\alpha}_{n(\tau)}(s):\mathbf{H}^{\alpha}(\tau)}$ where $\mathbf{H}^{\alpha}(\tau) = \mathbf{h}^{\alpha}(\tau) \otimes \mathbf{h}^{\alpha}(\tau)$. Their stored energy is then given by a Fung-type exponential stored energy function:
\begin{equation}
\hat{W}^{\alpha} (\mathbf{C}^\alpha_{n(\tau)} (s)) = \frac{c^{\alpha}_1}{4 c^{\alpha}_2}(\mathrm{exp}[c^{\alpha}_2(\mathbf{C}^{\alpha}_{n(\tau)}(s):\mathbf{H}^{\alpha}(\tau) - 1)^2] - 1).
\end{equation}
 Note that $c^\alpha$, $c^\alpha_1$, and $c^\alpha_2$ are constituent-specific material parameters typically determined from experimental data. Anisotropic constituents that exert active stress (such as smooth muscle cells) utilize the relationship
\begin{equation}
\boldsymbol{\sigma}^{act} (s) = \frac{\Phi^m_R (s)}{\text{det}(\mathbf{F}(s))} \mathbf{F} (s) \hat{\mathbf{S}}^{act} (s) \mathbf{F}^\text{T} (s),
\end{equation}
where $\Phi^m_R = \rho^m_R (s)/\hat{\rho}^m$ represents the referential volume fraction of constituent $m$ and
\begin{equation}
\hat{\mathbf{S}}^{act} (s) = T_{max} (1 - exp^{-C^2 (s)}) \lambda^{m(act)}_\theta (s) \left[ 1 - \left( \frac{\lambda_M - \lambda^{m(act)}_\theta}{\lambda_M - \lambda_0} \right)^2\right] \mathbf{H}^m_0.
\end{equation}
Here, $\mathbf{H}^m_0 = h^m_0 \otimes h^m_0$. $T_{max}$ is the maximum stress-generating capacity of the smooth muscle. $C(s)$ accounts for the ratio of vasoconstrictors to vasodilators through the relationship
\begin{equation}
C(s) = C_B - C_S \Delta \tau_f (s),
\end{equation}
where $C_B$ is the basal ratio and $C_S$ is a scaling factor that accounts for the effect deviations in wall shear stress have on vasoconstrictor behavior. $\lambda^{m(act)}_\theta$ is the circumferential stretch used to calculate active stress and is defined as
\begin{equation}
\lambda^{m(act)}_\theta = \int^s_{-\infty} k^{act} q^{act}(s,\tau) \left[ \mathbf{C}(\tau):\mathbf{H}^m_0 \right] \dd\tau,
\end{equation}
where we let
\begin{equation}
q^{act}(s,\tau) = e^{-k^{act}(s-\tau)}.
\end{equation}
This represents relaxation of the vasomotor tone over time.

\subsection{Constituent mass production and survival functions}
As previously defined in Section \ref{section:cmt}, the referential mass production function takes the form
\begin{equation}
m_R^\alpha(\tau) = m^\alpha_N \Upsilon^\alpha(\tau)
\end{equation}
where $\Upsilon^\alpha(\tau)$ is the constituent-specific stimulus function and the nominal production rate $m^\alpha_N (\tau)= \rho^\alpha_R(\tau)k^\alpha_h$ where $k^\alpha_h$ is related to the half-life of the constituent. For mechano-mediated constituent production, we utilize a linearized relation with a minimum value
\begin{equation} 
\Upsilon^\alpha_m(\tau) = 1 + K^\alpha_\sigma \Delta \sigma_{f} (\tau) - K^\alpha_{\tau} \Delta \tau_{f} (\tau) \quad\textrm{ for } \alpha = m
\end{equation}
\begin{equation} 
\Upsilon^\alpha(\tau) =  \begin{cases}
    0.1 & \Upsilon^\alpha_m(\tau) < 0.1  \\
    \Upsilon^\alpha_m(\tau) & \Upsilon^\alpha_m(\tau) \geq 0.1 
    \end{cases},
\end{equation}
where $ \Delta \sigma_{f} $ and $\Delta \tau_{f}$ represent deviations in intramural Cauchy stress and wall shear stress from homeostatic values respectively and are given by
\begin{equation}
\Delta \sigma_{f} (\tau)= \sigma_f (\tau)/\sigma_h - 1
\end{equation}
\begin{equation}
\Delta \tau_{f} (\tau)= \tau_f (\tau)/\tau_h - 1,
\end{equation}
and $K^\alpha_\sigma$ and $K^\alpha_\tau$ are gain-like parameters for production due to deviation from homeostatic values, $\sigma_h$ and $\tau_h$. $\sigma_f (\tau)$ is the radially averaged first-invariant of the intramural Cauchy stress while $\tau_f (\tau)$ is the wall shear stress at the inner radius. Both values are readily derived from the coupled fluid-solid solution.

One can add to the mechano-mediated stimulus a immuno-mediated stimulus. Such immunobiological factors can arise in a native artery in the case of excessive wall stress or damage; they can also arise as foreign body reactions in tissue engineering. For immuno-mediated constituent production in our tissue engineered graft simulation below, the constituent-specific stimulus function takes the form
\begin{equation}
\Upsilon^\alpha(\tau) = K^\alpha \frac{\Gamma^{alpha}(\tau)}{max(\Gamma^{alpha}(\tau))} \quad\textrm{ for } \alpha = i,
\end{equation}
with $\Gamma^{\alpha} = {\delta^\alpha}^{\beta^\alpha} \tau^{({\beta^\alpha} -1)} e^{-\delta^\alpha \tau}$ for $\alpha = i$, where $\delta$ and $\beta$ are rate and shape parameters of the inflammatory response and $K^\alpha$ is an inflammatory gain-like parameter.

Functional elastic fibers \textit{in-vivo} are produced only during the perinatal period, hence for both polymer ($\alpha = p$), in the case of implanted polymeric scaffolds for tissue engineering, and elastin ($e = p$) constituents
\begin{equation}
\Upsilon^\alpha(\tau) = 0 \quad\textrm{ for } \alpha = p,e.
\end{equation}

Survival functions for both mechano- and immuno-mediated productions take the form
\begin{equation}
q^\alpha(s,\tau) = exp\left(-\int^s_\tau k^\alpha(t) dt \right) \quad\textrm{ for } \alpha = i,m.
\end{equation}
 where $k^\alpha(t) = k^a_h( 1 + K^\alpha_{D_\sigma} \Delta \sigma_f (t) + K^\alpha_{D_\tau} \Delta \tau_f (t))$ for $\alpha = m$ and $k^\alpha(t) = k^a_h( 1 + K^\alpha(t)/K^\alpha_{max})$ for $\alpha = i$.

For elastin in homeostasis
\begin{equation}q^\alpha (s, \tau) = 1 \quad\textrm{ for } \alpha = e\end{equation}
as the mass fraction of elastin stays approximately constant \textit{in-vivo} after maturation, except in diseases characterized by heightened elastolysis. In mechanically mediated constituents, at homeostasis $\Delta \tau_f (s)$ and $\Delta \sigma_f (s)$ are presumed to be $0$ and the rate of mass degradation is presumed to equal the rate of mass production. This yields the relationship 
\begin{equation}
{m^\alpha_N}_h = \rho^\alpha_R(0)k^\alpha_h \quad\textrm{ for } \alpha = m,
\end{equation}
which represents the homeostatic mass production rate. To account for cases where there is initially no existing mechano-mediated constituent (such as in tissue-engineered grafts) we set the basal production rate as
\begin{equation}
m^\alpha_N (\tau) = \text{max}\left([\rho^\alpha_R(\tau)k^\alpha_h,{m^m_N}_h] \right) \quad\textrm{ for } \alpha = m,
\end{equation}
ensuring that there is always a minimum reference mass. Additionally, in the case of inflammation-mediated constituents
\begin{equation}
m^\alpha_N (\tau) = {m^m_N}_h \textrm{ for } \alpha = i,
\end{equation}
meaning that the nominal production rate is determined by the constituent's mechano-mediated counterpart. The survival function of polymer constituents take the form of
\begin{equation}
Q^\alpha (s) = (1 + exp(-k^\alpha \zeta^\alpha)) / (1 + exp(k^\alpha \gamma^\alpha (s - \zeta^\alpha / \gamma^\alpha))) \quad\textrm{ for } \alpha = p,
\end{equation}
which represents their \textit{in-vivo} degradation. If the polymer is part of a defined ground matrix it instead has the survival function
\begin{equation}
Q^\alpha (s) = \begin{cases} 
      1 & s < s^{a}_{d} \\
      ((1 - \varepsilon^{a}_{Rmin}) e^{(-k^{\alpha}_h (s - s^{a}_{d})) + \varepsilon^{a}_{Rmin})} & s^{a}_{d} \leq s \\
   \end{cases} \quad\textrm{ for } \alpha = p_{gnd}.
\end{equation}

Together, these mass turnover and survival functions specify the behavior of all constituents utilized to model vascular behavior in Section \ref{sec:illustrative}.

\subsection{Simulation boundary conditions}

In the solid and fluid formulations we apply boundary conditions consistent with known values within the vascular system. Given a known volumetric flow rate profile $Q(\iota)$, we define $\Gamma^f_h$, the essential boundary condition on the fluid domain, as our inlet with the flow rate mapped to a parabolic velocity profile implemented on the domain as
\begin{equation}
\mathbf{u}(\iota, \mathbf{x}) = Q(\iota) \frac{|\mathbf{x}_i - \mathbf{x}_c|^2}{\sum |\mathbf{x}_i - \mathbf{x}_c|^2} \text{ on } \Gamma^f_h,
\end{equation}
where $\mathbf{x}_c$ is the geometric center of the inflow face. We define a resistance-based Neumman boundary at the outlets of the fluid domains such that
\begin{equation}
p = R \int_{\Gamma^f_g} \mathbf{u} \text{ on } \Gamma^f_g.
\end{equation}
On the structural conditions, the solution from the fluid formulation provides the traction boundary condition at the interface. To mimic physiological conditions, the vessels are modeled as axially clamped such that
\begin{equation}
\mathbf{d} \cdot \hat{\mathbf{n}} = 0 \text{ on } \Gamma^s_d,
\end{equation}
where $\Gamma^s_d$ are the inlet and outlet face of the solid domain. To prevent spurious rigid body motions, on patient-specific vascular domains we apply a spring-like boundary condition on the outer surface of the vascular wall
\begin{equation}
\mathbf{f} = -k_s \mathbf{d} \text{ on } \Gamma^s_w,
\end{equation}
where $\Gamma^s_w$ is the outer wall of the solid domain. We choose $k_s$ to have a value significantly lower than the other characteristic forces in the problem so that it does not have a significant impact on the solution fields.

\section{Illustrative Examples}\label{sec:illustrative}

In this section, we will demonstrate the utility of the FSG formulation. First, we validate the FSG framework on idealized veins and show that it agrees with previous, reduced-order implementations of CMT. Next, we demonstrate the impact that inclusion of the full fluid domain has on 3D outcomes by comparing the outcomes of G\&R on venous tissue-engineered vascular grafts (TEVGs). Then, we demonstrate the ability of the FSG formulation to simulate large volume changes in patient-specific geometries by simulating TEVG behavior on an ovine graft segmented directly from clinical imaging. We further show the framework's robustness to vascular type, timescale, and pathological condition by simulating the remodeling of a patient-specific human aorta under both normal and hypertensive conditions. We finally demonstrate the ability of this framework to simulate pulsatile conditions by comparing pulsatile behavior of the normal and hypertensive aorta.

\begin{table}[!h]
\begin{center}
\begin{tabular}{ c c c } 
 \hline
 \hline

\multicolumn{3}{c}{\textbf{Ovine IVC Properties}} \\ 
 \hline
 \hline
 Initial volume fractions  & $\Phi^e_0,\Phi^m_0,\Phi^c_0$ & 0.1,  0.081, 0.8189 \\ 
 \hline
 True mass densities  & $\hat{\rho}^{e},\hat{\rho}^{m},\hat{\rho}^{c}$ & 1050, 1050, 1050 kg/m\textsuperscript{3} \\ 
 \hline
 Collagen fractions & $\beta^\theta, \beta^z, \beta^d$ & 0.0139, 0.137, 0.668 \\
 \hline
Diagonal collagen orientation & $\alpha_0$ & $\pm 41.94$\textdegree  \\ 
 \hline
Elastin parameter & $c^e$ & 9.913 kPa\\ 
 \hline
Smooth muscle parameters & $c^m_1, c^m_2$ & 48.330 kPa, 1.02  \\ 
 \hline
Collagen parameters & $c^c_1, c^c_2$ & 2.696 kPa, 14.92  \\ 
 \hline
Elastin deposition stretches & $G^e_r,G^e_\theta,G^e_z$ & 1/($G^e_\theta G^e_z$), 1.219, 1.428  \\ 
 \hline
Constituent removal rate parameter  & $k^e_h,k^m_h,k^c_h$ & -, 1/80, 1/80 days  \\ 
 \hline
WSS mediated production gains & $K^e_{WSS},K^m_{WSS},K^c_{WSS}$ & -, 5.0, 5.0  \\ 
 \hline
Intramural stress mediated production gains & $K^e_\sigma,K^m_\sigma,K^c_\sigma$ & -,1.0,1.0  \\
\hline
WSS mediated degradation gains & $K^e_{D_{WSS}},K^m_{D_{WSS}},K^c_{D_{WSS}}$ & -,0,0  \\ 
 \hline
Intramural stress mediated degradation gains & $K^e_{D_\sigma},K^m_{D_\sigma},K^c_{D_\sigma}$ & -,0,0  \\ 
 \hline
Homeostatic WSS & $\tau_h$ &  0.16166 Pa \\  
 \hline
Homeostatic intramural stress & $\sigma_h$ &  0.615 kPa \\  
 \hline
Maximum active stress & $T_{max}$ &  - kPa\\ 
 \hline
Active stress remodeling time & $k_{act}$ &  - days \\ 
 \hline
Basel vasoconstriction, Vasoconstriction scaling & $C_B$, $C_S$ &  -, -\\
 \hline
\end{tabular}
\end{center}
\caption{Vessel simulation parameters of an ovine IVC informed by the parameters described previously.\cite{latorre2022vivo}}
\label{tab:venous}
\end{table}

\begin{table}[!h]
\begin{center}
\begin{tabular}{ c c c } 
 \hline
 \hline

\multicolumn{3}{c}{\textbf{Tissue-Engineered Neovessel Properties}} \\ 
 \hline
 \hline
Initial volume fractions  & $\Phi^{p1}_0,\Phi^{p2}_0,\Phi^{gnd}_0$ &  0.15, 0.05, 0.8 \\ 
 \hline
True mass densities  & $\hat{\rho}^{p1}$,$\hat{\rho}^{p2}$,$\hat{\rho}^{gnd}$ &  1230, 1530, 1050 kg/m\textsuperscript{3} \\ 
 \hline
Material parameters & $c^1$, $c^2$ & 2000000, 1000 kPa  \\ 
 \hline
Polymer degradation parameters & $k^{p1}, k^{p2}, k^{gnd}$ & 1/80, 1/10, 1/80 days\\ 
  & $\zeta^1,\zeta^2$ & 245, 60\\ 
  & $\gamma^{p1},\gamma^{p2}$ & 4, 4\\ 
  & $\varepsilon^{gnd}_{Rmin}$ & 0.10\\ 
  &  $s^{gnd}_d$ & 14 days\\ 
 \hline
Inflammatory collagen parameters & $c^i_1, c^i_2$ & 80.88 kPa, 14.92  \\
\hline
Inflammatory tissue production parameters & $K^i$, $\delta^i$, $\beta^i$ & 26.8196, 0.25, 5.0 \\
 \hline
Inflammatory tissue degradation parameter & $k^i$ & 0.3576 \\
 \hline
\end{tabular}
\end{center}
\caption{Vessel simulation parameters of a polymer-based TEVG informed by parameters described previously.\cite{szafron2018immuno,latorre2022vivo}}
\label{tab:tevg}
\end{table}

\subsection{Validation of finite element framework}

\begin{figure}[!h]
\includegraphics[width=\textwidth]{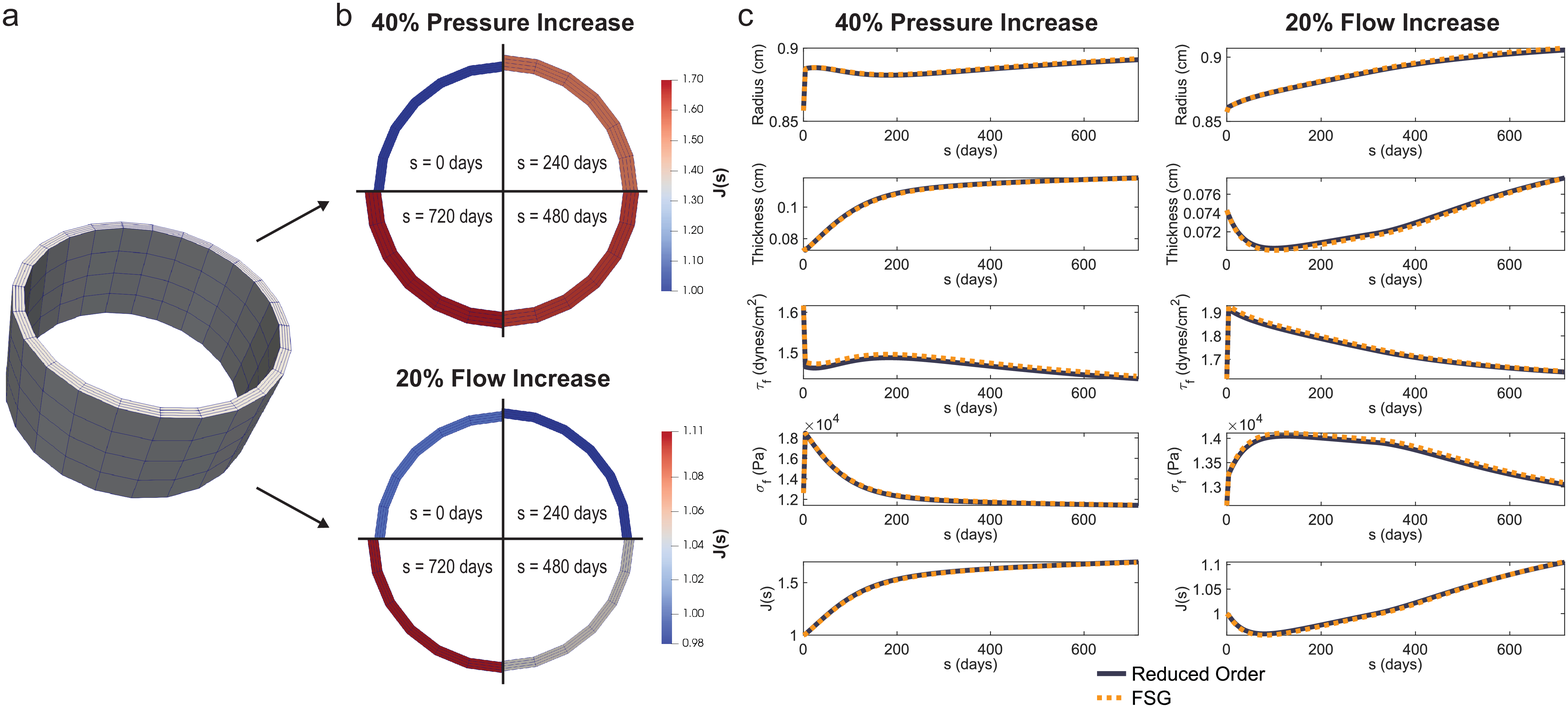}
\centering
\caption{Initial (a) cylindrical geometry of an ovine IVC. Increased pressure and flow (b) simulations were compared to results using a reduced-order constrained mixture model (c). The finite element results are shown by the black, solid line while the reduced-order results are shown by the purple, dotted line.}
\label{fig:validation}
\end{figure}

To validate our finite-element implementation, we conducted a comparative analysis between our framework and a previously utilized framework which solved evolving configurations of a thin-walled, axisymmetrical vessel slice. \cite{drews2020spontaneous} This comparison was performed using the vessel properties of the ovine inferior vena cava (IVC) with initial radius of $0.8573$ cm and initial thickness of $0.743$ cm informed by the parameters described in Latorre et al. 2022 and outlined in Table \ref{tab:venous}.\cite{latorre2022vivo} The finite element model had an initial length of length of $0.8573$ cm where the reduced order model was considered of infinite length. A G\&R timestep size of $4$ days was used and the simulations were run for a total duration of $720$ days. The convergence criteria of the finite-element framework was set to be $\epsilon_r = 1\mathrm{e}{-5}$ on the displacement residual field and $1\mathrm{e}{-4}$ on all other fields. In the hypertensive case, an inflow of $Q = 20$ ml/s was applied to both the FSG and reduced order model. A constant inner wall pressure of $8610$ $\frac{\text{dynes}}{\text{cm}^2}$ was applied to the hypertensive reduced order model while an outlet resistance of $R = 430.5$ $\frac{\text{dynes}\cdot \text{s}}{\text{cm}^5}$ was applied to the FSG model. This represents a 40\% increase in pressure from the homeostatic value. In the increased flow case, an steady inflow of $Q(\iota) = 24$ ml/s was applied to both the FSG and reduced order model. A constant inner wall pressure of $6140$ $\frac{\text{dynes}}{\text{cm}^2}$ was applied to the hypertensive reduced order model while an outlet resistance of $R = 307.5$ $\frac{\text{dynes}\cdot \text{s}}{\text{cm}^5}$ was applied to the FSG model. This represents a 20\% increase in flow from the homeostatic value. Algorithm \ref{fsg_coupling} was used for both simulations.

There is excellent agreement between our finite-element framework and the previously validated reduced-order framework (Figure \ref{fig:validation}). This shows that the FSG framework recapitulates time-resolved solutions for known classes of G\&R problems.

\subsection{Idealized tissue-engineered vascular graft}

\begin{figure}[!h]
\includegraphics[width=\textwidth]{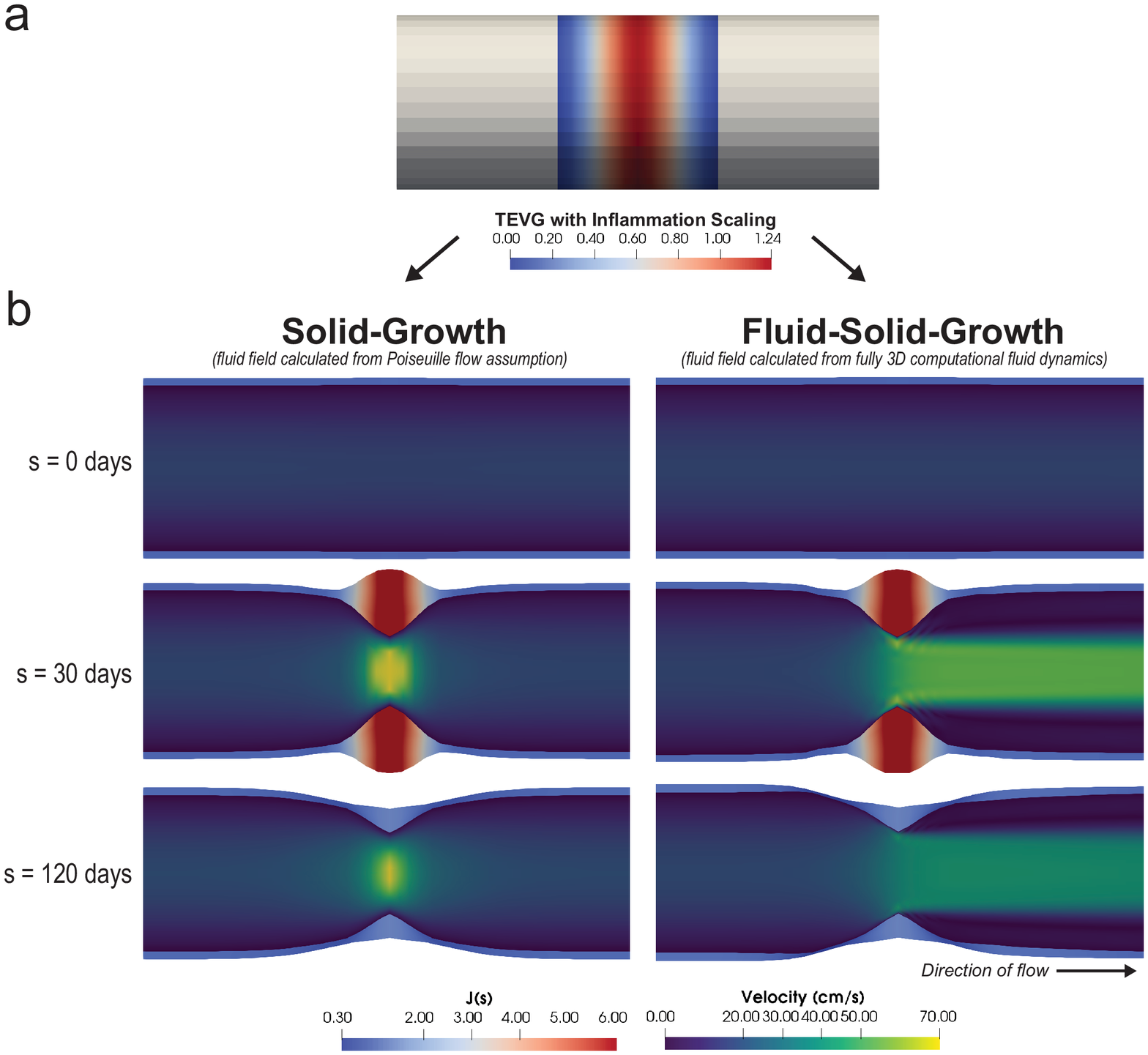}
\centering
\caption{Impact of fluid field on outcomes are shown by simulation of an interpositional TEVG (a) with both reduced order hemodynamics (b, left) and full fluid field computation (b, right).}
\label{fig:fluidimpact}
\end{figure}

\begin{figure}[!h]
\includegraphics[width=\textwidth]{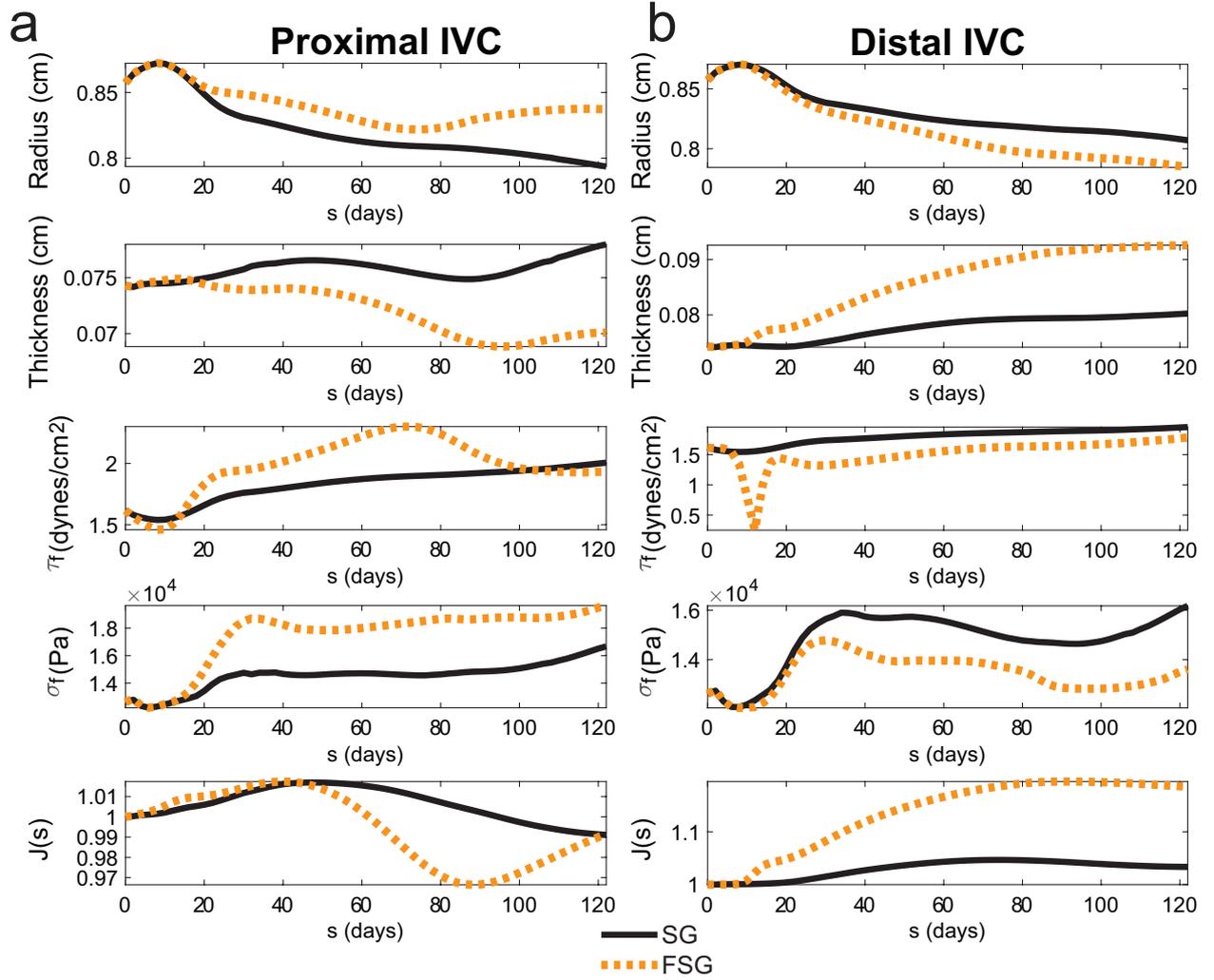}
\centering
\caption{Inclusion of a fully 3D fluid field in a ovine TEVG simulation results in markedly divergent remodeling both proximally (a) and distally to the central stenosis (b).}
\label{fig:fluidimpact_proxdist}
\end{figure}

\begin{figure}[!h]
\includegraphics[width=\textwidth]{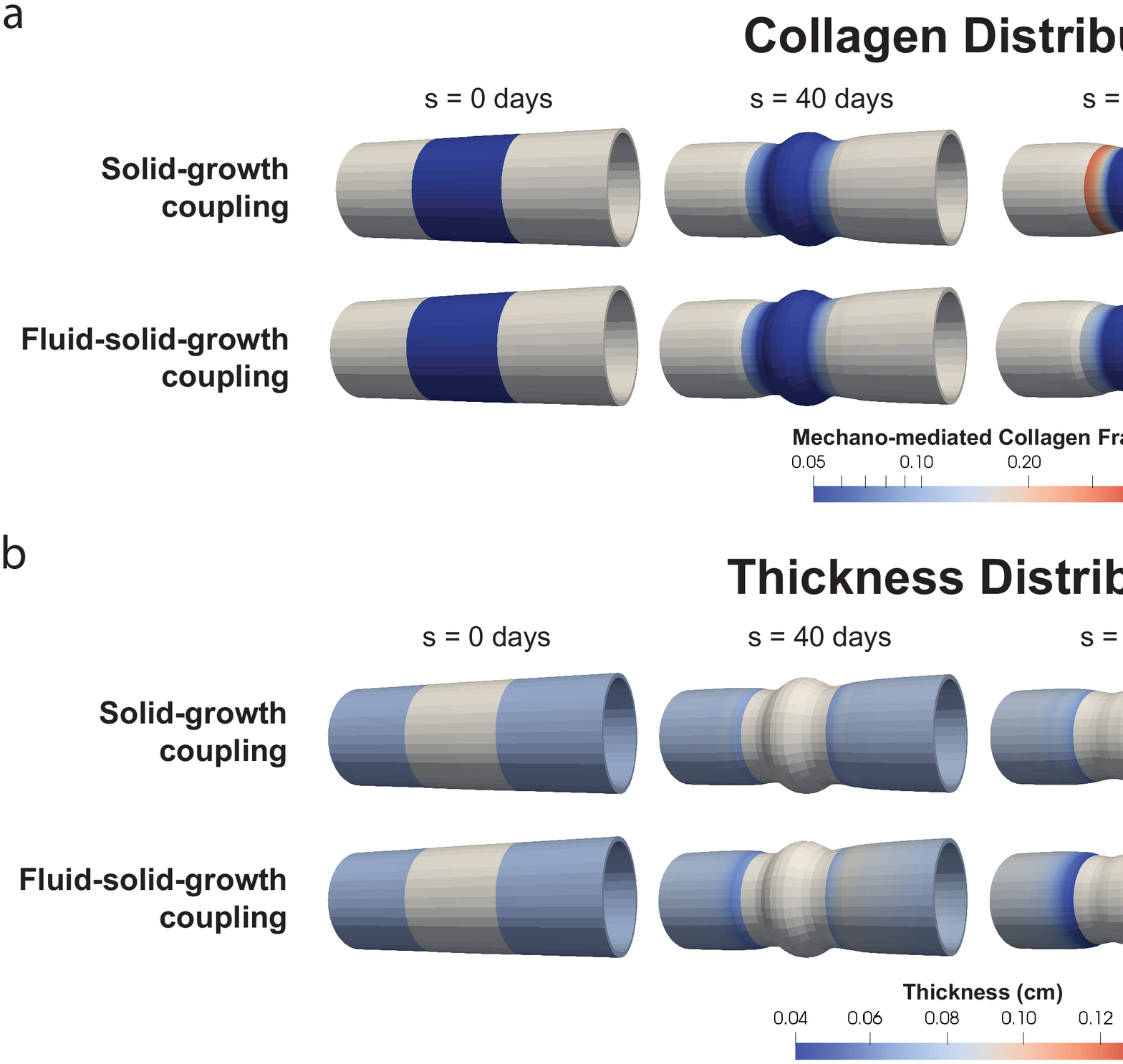}
\centering
\caption{Difference between mechano-mediated collagen distribution within the TEVG (a) and the thickness of native vasculature attached to the TEVG (b).}
\label{fig:fluidimpact_distribution}
\end{figure}

We show the importance of considering a fully resolved fluid field when describing the evolution of vascular geometries by comparing here the outcomes of two tissue-engineered vascular graft simulations motivated by parameters outlined previously.\cite{latorre2020fast,szafron2018immuno} In these simulations, a TEVG is attached between two sections of native IVC vasculature, such as in an interpositional IVC graft. Both of these simulations utilize identical parameters described in Table \ref{tab:venous} and Table \ref{tab:tevg}. In the first case, we utilize a purely solid-growth formulation. Here, a constant pressure condition is assumed throughout the G\&R simulation and does not evolve with the vasculature. The wall shear stress is assumed to follow the radius-based Poiseuille assumption. These assumptions are standard practice in many constrained mixture frameworks that seek to simplify the calculation of fluid-mediated inputs. In the second case, we utilize our full FSG framework whereby the Navier-Stokes equations are solved alongside the solid-growth equations which creates a heterogeneous flow field that evolves with the vessel. We begin, in both cases, with a cylinder of initial radius of $0.8573$ cm, initial thickness of $0.743$ cm, and an initial length of $3.4292$ cm. The tissue-engineered grafts is modeled to be interposed between two segments of inferior vena cava vasculature (Figure \ref{fig:validation}a). In both simulations, a G\&R timestep size of $2$ days was used and the simulations were run for a total duration of $120$ days. A flow rate of $Q(\iota) = 20$ ml/s is imposed along with an outlet resistance of $R = 307.5$ $\frac{\text{dynes}\cdot \text{s}}{\text{cm}^5}$ which results in an outlet pressure of $6150$ $\frac{\text{dynes}}{\text{cm}^2}$. In the solid-growth simulation, we assume a constant internal pressure of $6150$ $\frac{\text{dynes}}{\text{cm}^2}$ . The convergence criteria of the finite-element framework was set to be $\epsilon_r = 1\mathrm{e}{-3}$ on the displacement residual field and $1\mathrm{e}{-2}$ on all other residual fields. Algorithm \ref{fsg_coupling} was used in both simulations, with the fluid field in the purely solid-growth coupling calculated using the assumptions noted above.

TEVGs are marked by inflammatory mass recruitment that can cause mild to severe stenosis. This is displayed in both simulated grafts (Figure \ref{fig:fluidimpact}b). However, only the full FSG formulation incorporates the effects that the stenosis has on the fluid field (Figure \ref{fig:fluidimpact_proxdist}). In the vasculature proximal to the stenosis, the increased pressure induced by the resistance of the stenotic region results in increased intramural stress and an increased radius, which lowers the experienced wall shear stress. Both these changes contribute to slight volume recruitment and thickening (Figure \ref{fig:fluidimpact_proxdist} and Figure \ref{fig:fluidimpact_distribution}a). However, as maximal stenosis is reached, the vascular wall proximal to the stenosis becomes more perpendicular to the direction of flow which increases the experienced wall shear stress. This results in volume loss and eventual thinning of the proximal vessel wall (Figure \ref{fig:fluidimpact_proxdist} and Figure \ref{fig:fluidimpact_distribution}a). By contrast, the distal IVC experiences a marked drop in wall shear stress due to the formation of a velocity jet through the stenosis. This results in prolonged mass recruitment and thickening that persists throughout the simulation's duration (Figure \ref{fig:fluidimpact}c and Figure \ref{fig:fluidimpact_distribution}a). Initially, the effects of the fluid field have relatively little impact on the TEVG portion of the simulation as the immuno-mediated constituent recruitment is not mechano-sensitive. However, at later stages when mechano-mediated constituents are increasingly recruited, we see that the proximal portion of the graft is expected to have a smaller portion of mechano-mediated constituents due to the high shear stress experienced by the proximal graft wall (Figure \ref{fig:fluidimpact_proxdist} and Figure \ref{fig:fluidimpact_distribution}b).

This comparison highlights importance of using full resolution fluid fields in complex-domain CMT applications and shows that remodeling in one portion of a vessel can have a profound effect on the hemodynamic stresses experienced by the surrounding vasculature. To calculate metrics of time-resolved vascular performance, such as overall vascular resistance, the full FSG interaction is needed.

\subsection{Patient-specific tissue-engineered vascular graft}

\begin{figure}[!h]
\includegraphics[width=0.9\textwidth]{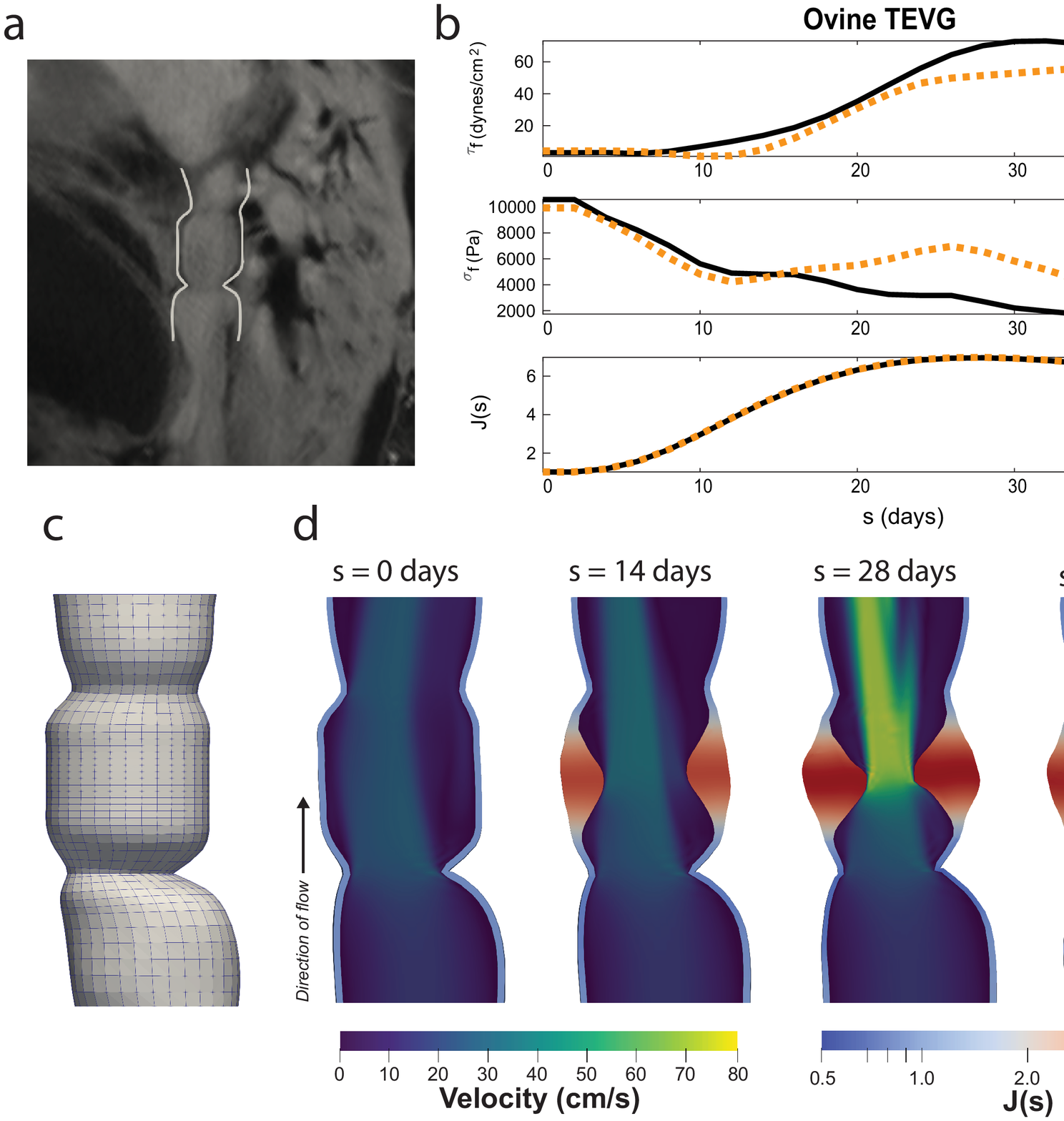}
\centering
\caption{We segment a subject specific geometry from clinical imaging (a) and create a finite element model (b). We simulate growth and remodeling for 6 weeks (c,d) and demonstrate that the subject-specific geometry mediates growth and remodeling results (c,d).}
\label{fig:ps_tevg}
\end{figure}

\begin{figure}[!h]
\includegraphics[width=0.9\textwidth]{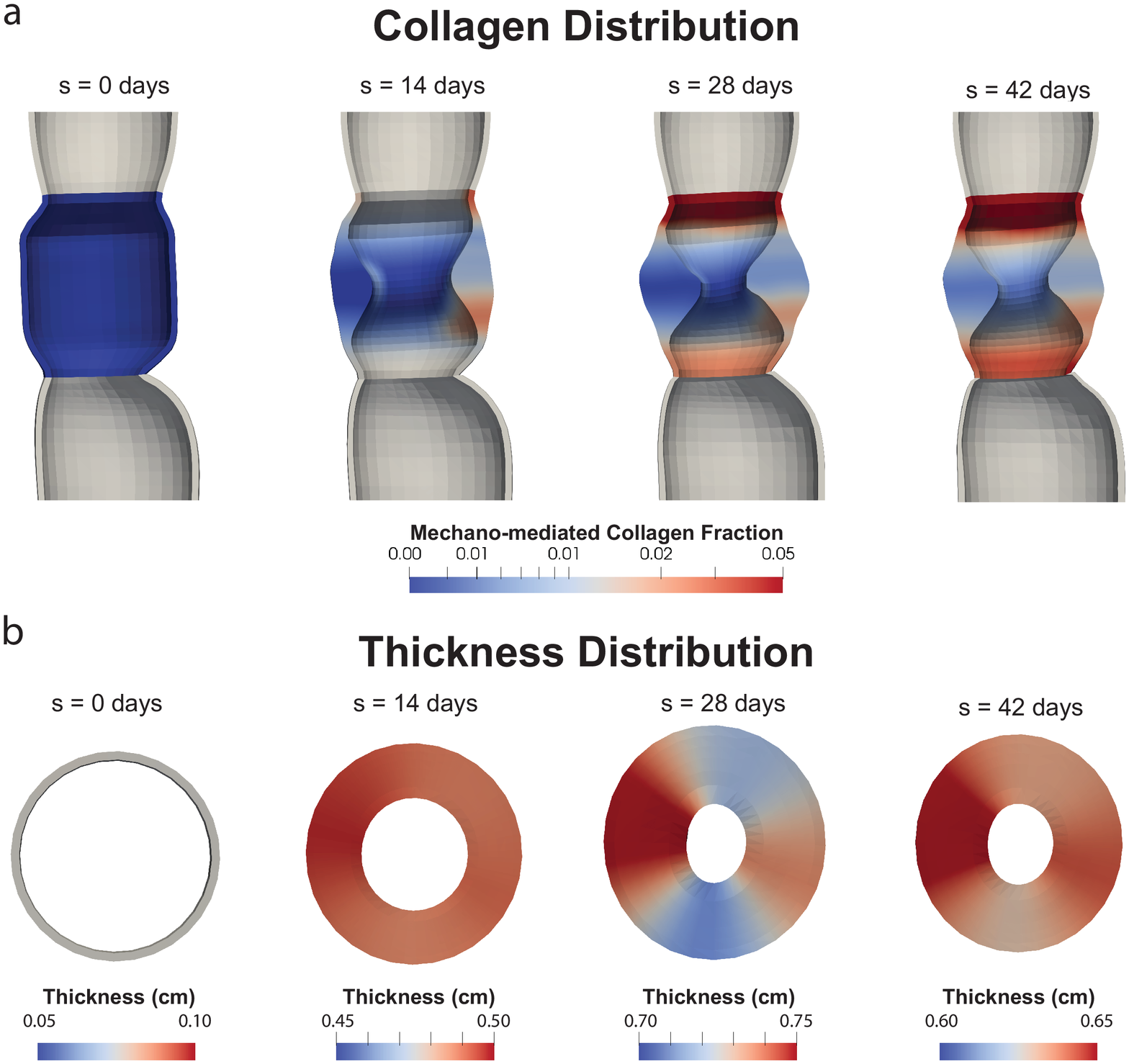}
\centering
\caption{Frontal cut-plane view of TEVG model showing distribution of mechano-mediated collagen (a) and axial cut-plane view of the mid-TEVG thickness in a patient-specific TEVG.}
\label{fig:distribution_ps_tevg}
\end{figure}

We then expanded this framework to a patient-specific geometry segmented directly from clinical imaging. Here, we model a tissue-engineered vascular graft in a sheep implanted interpositionally between two segments of the IVC (Figure \ref{fig:ps_tevg}) and imaged at 1 week post-TEVG implantation. Details on graft composition and surgical procedure can be found in elsewhere.\cite{blum2022tissue} MRI images were taken at 1 week post-implantation. A G\&R timestep size of $2$ days was used and the simulation was run for a total duration of $42$ days. A flow rate of $Q(\iota) = 20$ ml/s was imposed along with an outlet resistance of $R = 307.5$ $\frac{\text{dynes} \cdot \text{s}}{\text{cm}^5}$ for an outlet pressure of $p = 6150$ $\frac{\text{dynes}}{\text{cm}^2}$. The convergence criteria of the finite-element framework was set to be $\epsilon_r = 1\mathrm{e}{-3}$ on the displacement residual field and $1\mathrm{e}{-2}$ on all other residual fields. A spring constant of $k_s = 100$ $\frac{\text{dynes}}{\text{cm}^3}$ was applied to the outer wall of the vessel. Algorithm \ref{alt_fsg_coupling} was used for FSG coupling.

The subject-specific geometry mediated the fluid field throughout the entirety of the simulation (Figure \ref{fig:ps_tevg}b). Generally, it produced a higher wall shear stress on the left-lateral portion of the TEVG wall(Figure \ref{fig:ps_tevg}c); this created a preferential deposition of mechano-mediated constituents on the right-lateral portion of the TEVG wall (Figure \ref{fig:distribution_ps_tevg}). This finding demonstrates that subject-specific geometry mediates the ability of the model to successfully incorporate mechano-mediated constituents, which will later determine the TEVG's ability to adapt to its hemodynamic environment.  In addition, the the graft geometry is distorted from its original axisymmetrical shape throughout its growth and remodeling. By six weeks, this results in thickness that varies by more than 10\% at mid-TEVG and has an elliptical geometry with a minimum diameter of 0.58 cm and maximum diameter of 0.74 cm (Figure \ref{fig:distribution_ps_tevg}). This recapitulates clinical observations of ovine TEVGs where diameter decreased by over 50\% and evolved into an elliptical geometry at six weeks post-implantation.\cite{drews2020spontaneous,blum2022tissue}

\subsection{Patient-specific aortic arch}

\begin{table}[!h]
\begin{center}
\begin{tabular}{ c c c } 
 \hline
 \hline
\multicolumn{3}{c}{\textbf{Human Aortic Properties}} \\ 
 \hline
 \hline
  Initial volume fractions  & $\Phi^e_0,\Phi^m_0,\Phi^c_0$ & 0.1,  0.0604, 0.8396 \\ 
 \hline
 True mass densities  & $\hat{\rho}^{e},\hat{\rho}^{m},\hat{\rho}^{c}$ & 1050, 1050, 1050 kg/m\textsuperscript{3} \\ 
 \hline
 Collagen fractions & $\beta^\theta, \beta^z, \beta^d$ & 0.0, 0.067, 0.7792 \\
 \hline
Diagonal collagen orientation & $\alpha_0$ & $\pm45$\textdegree  \\ 
 \hline
Elastin parameter & $c^e$ & 553.287 kPa\\ 
 \hline
Smooth muscle parameters & $c^m_1, c^m_2$ & 784.3185 kPa, 29.16  \\ 
 \hline
Collagen parameters & $c^c_1, c^c_2$ & 784.3185 kPa, 29.16  \\ 
 \hline
Elastin deposition stretches & $G^e_r,G^e_\theta,G^e_z$ & 1/($G^e_\theta G^e_z$), 1.257, 1.238  \\ 
 \hline
Constituent removal rate parameter  & $k^e_h,k^m_h,k^c_h$ & -, 1/70, 1/70 days  \\ 
 \hline
WSS mediated production gains & $K^e_{WSS},K^m_{WSS},K^c_{WSS}$ & -, 1.0, 1.0  \\ 
 \hline
Intramural stress mediated production gains & $K^e_\sigma,K^m_\sigma,K^c_\sigma$ & -,1.0,1.0  \\
\hline
WSS mediated degradation gains & $K^e_{D_{WSS}},K^m_{D_{WSS}},K^c_{D_{WSS}}$ & -,0,0  \\ 
 \hline
Intramural stress mediated degradation gains & $K^e_{D_\sigma},K^m_{D_\sigma},K^c_{D_\sigma}$ & -,0,0  \\ 
 \hline
Homeostatic WSS & $\tau_h$ &  0.494 Pa \\  
 \hline
Homeostatic intramural stress & $\sigma_h$ &  13.33 kPa \\  
 \hline
Maximum active stress & $T_{max}$ &  39.86 kPa\\ 
 \hline
Active stress remodeling time & $k_{act}$ &  1/70 days \\ 
 \hline
Basel vasoconstriction, Vasoconstriction scaling & $C_B$, $C_S$ &  0.7, 1.2\\
 \hline
\end{tabular}
\end{center}
\caption{Vessel simulation parameters of a human aorta informed by parameters described previously.\cite{figueroa2009computational}}
\label{tab:arterial}
\end{table}

\begin{figure}[!h]
\includegraphics[height=0.9\textheight]{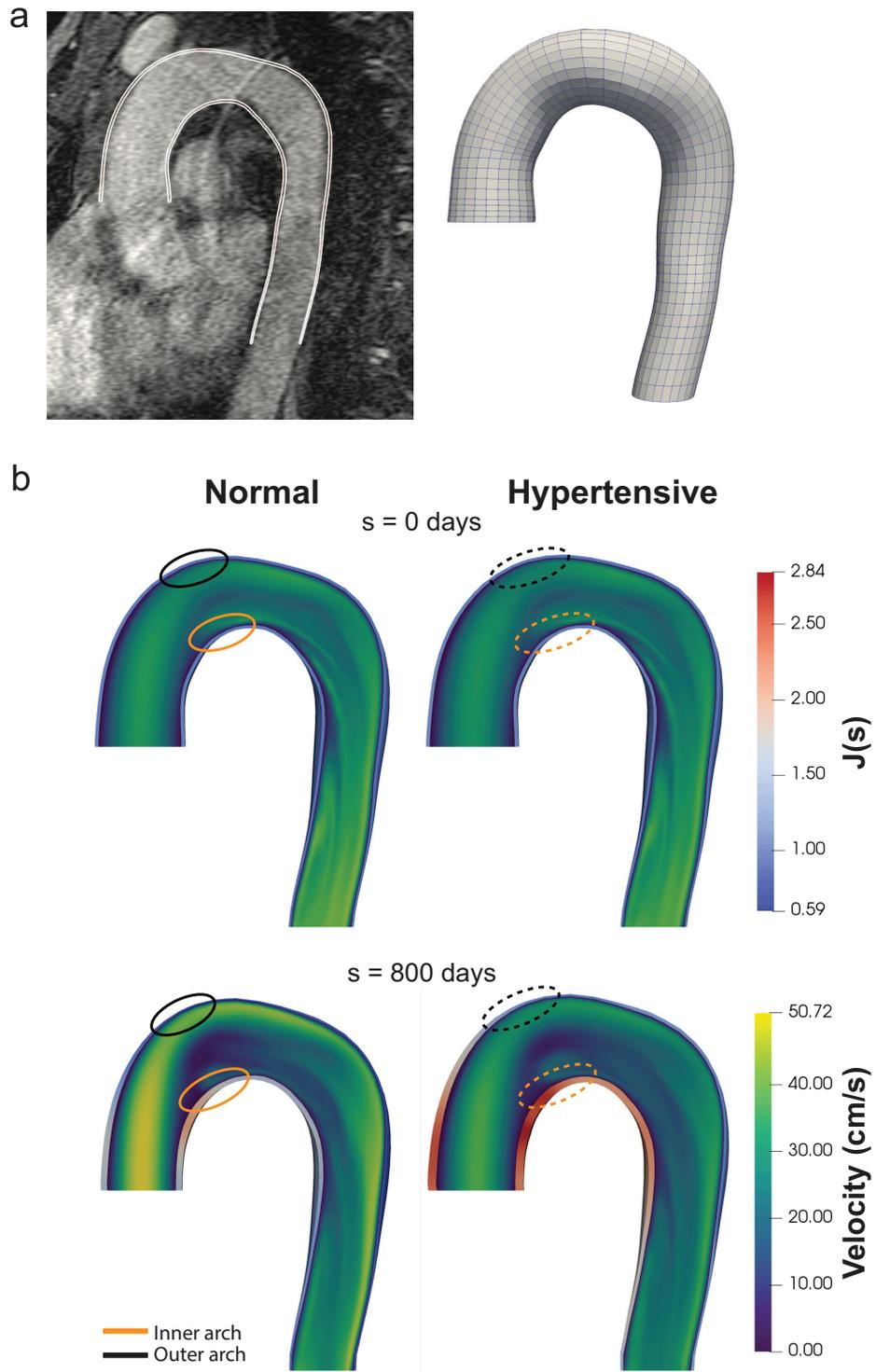}
\centering
\caption{Aorta model generated from patient-specific imaging (a) under normal (b, left) and hypertensive (b, right) conditions.}
\label{fig:ps_aorta}
\end{figure}

\begin{figure}[!h]
\includegraphics[width=\textwidth]{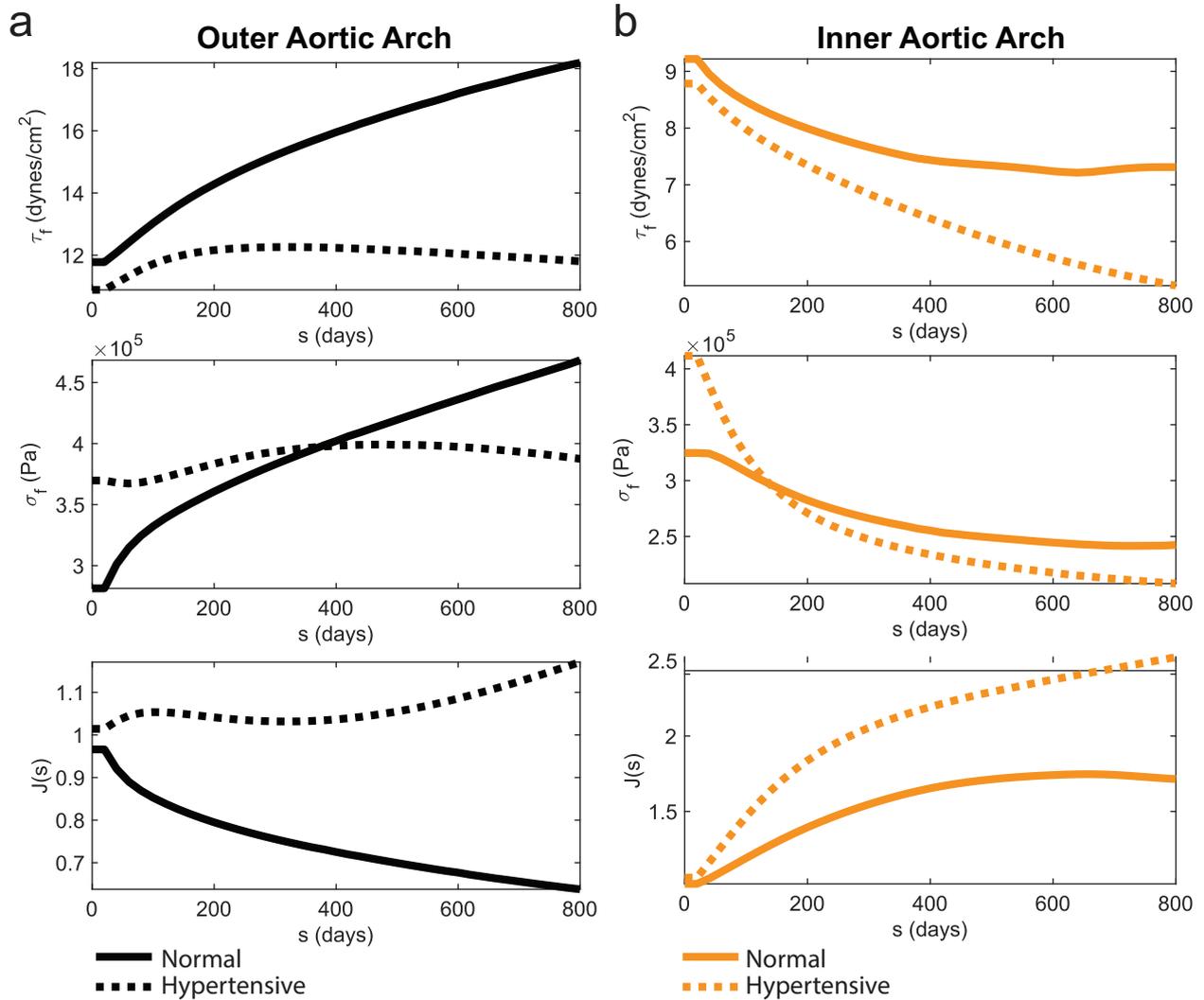}
\centering
\caption{In patient-specific aortas, there are significant differences in the time-resolved remodeling of the outer (a) and inner (b) aortic arch that are both geometry and pathology-dependent.}
\label{fig:ps_aorta_inout}
\end{figure}

\begin{figure}[!h]
\includegraphics[width=\textwidth]{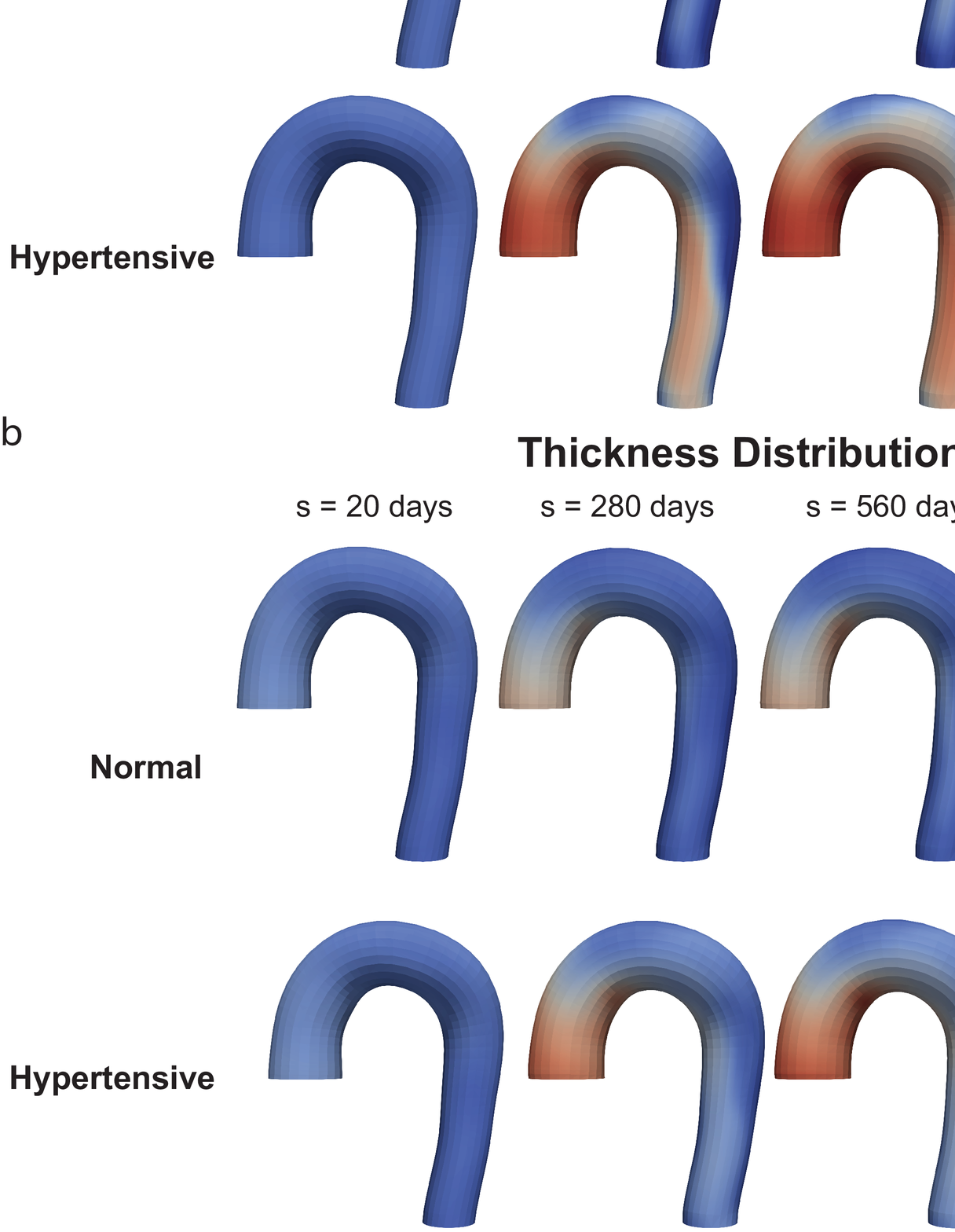}
\centering
\caption{Distribution of collagen (a) and thickness (b) in a normal and hypertensive aorta model.}
\label{fig:ps_aorta_distribution}
\end{figure}

To illustrate the ability of this framework to be used in arterial settings, we use a patient-specific aortic arch segmented from the MRI images of a healthy 27 year old from the Vascular Model Repository (https://www.vascularmodel.com/) and arterial parameters scaled from those found elsewhere (Table \ref{tab:arterial}).\cite{figueroa2009computational} A G\&R timestep size of $20$ days was used and the simulation was run for a total duration of $800$ days. A flow rate of $Q = 97.0$ ml/s was imposed along with an outlet resistance of $R = 1374.53$ $\frac{\text{dynes} \cdot \text{s}}{\text{cm}^5}$ for an outlet pressure of $p = 13333.33$ $\frac{\text{dynes}}{\text{cm}^2}$. The convergence criteria of the finite-element framework was set to be $\epsilon_r = 1\mathrm{e}{-3}$ on the displacement residual field and $1\mathrm{e}{-2}$ on all other residual fields. A spring constant of $k_s = 100000$ $\frac{\text{dynes}}{\text{cm}^3}$ was applied to the outer wall of the vessel. Algorithm \ref{alt_fsg_coupling} was used for FSG coupling.

The model was not at homeostasis at the initial time as the arterial parameters were optimized for an idealized tube geometry which did not consider the impact of the patient-specific geometry and fluid flow field. Because of this, initial remodeling causes thickening at the aortic root and on the inner curvature of the aortic arch due to low wall shear stress (Figure \ref{fig:ps_aorta} and Figure \ref{fig:ps_aorta_inout}). The outer curvature of the aortic arch experiences comparatively high wall shear stress and thins in response. This illustrative example does not account for many features of a full aortic model, notably it lacks superior aortic branches that would siphon flow from the descending aorta, thus reducing wall shear stress and pressure along the descending aorta. However, it successfully illustrates the applicability of the finite element formulation to patient-specific geometries at a human scale. It also serves as a basis for comparison to pathological hemodynamic conditions.

To compare to hypertensive conditions, we model the same aortic geometry but apply a distal resistance of $1924.35$ $\frac{\text{dynes} \cdot \text{s}}{\text{cm}^5}$, representing a 40\% pressure increase from the originally modeled case. Under these conditions, there is mass production and thickening in response to the increased pressure conditions, as expected. However, the rate and magnitude of thickness increase varies across the geometry and does not scale solely based on pressure increase (Figure \ref{fig:ps_aorta}, Figure \ref{fig:ps_aorta_inout}, and Figure \ref{fig:ps_aorta_distribution}).  In the outer curvature of the hypertensive ascending aorta, the instantaneous pressure increase causes both increased intramural stress and decreased wall shear stress due to both dilation of the ascending aorta and also pressure-induced dilation at the aortic root which causes overall decreased velocity as flow rate is preserved. While the wall shear stress is still elevated compared to the homeostatic target ($12$ $\frac{\text{dynes}}{\text{cm}^2}$ vs $10$ $\frac{\text{dynes}}{\text{cm}^2}$), the reduction in wall shear stress allows the intramural stress gain function to initially dominate the behavior of mass recruitment, increasing the overall volume. This, in turn, causes the wall shear stress to increase slightly, but at intermediate to late times, the ascending aorta generally increases in radius, once again lowering wall shear stress and allowing the elevated intramural stress to dominate mass recruitment. In the inner curvature of the aortic arch, a similar phenomenon happens, although persistent depression of the wall shear stress drives mass recruitment well past the point at which homeostatic intramural stress is recovered. This behavior demonstrates that the outcomes of pathological conditions are also highly mediated by patient-specific geometry and full resolution hemodynamic fields.

\subsection{Simulation under pulsatile conditions}

\begin{figure}[!h]
\includegraphics[height=0.9\textheight]{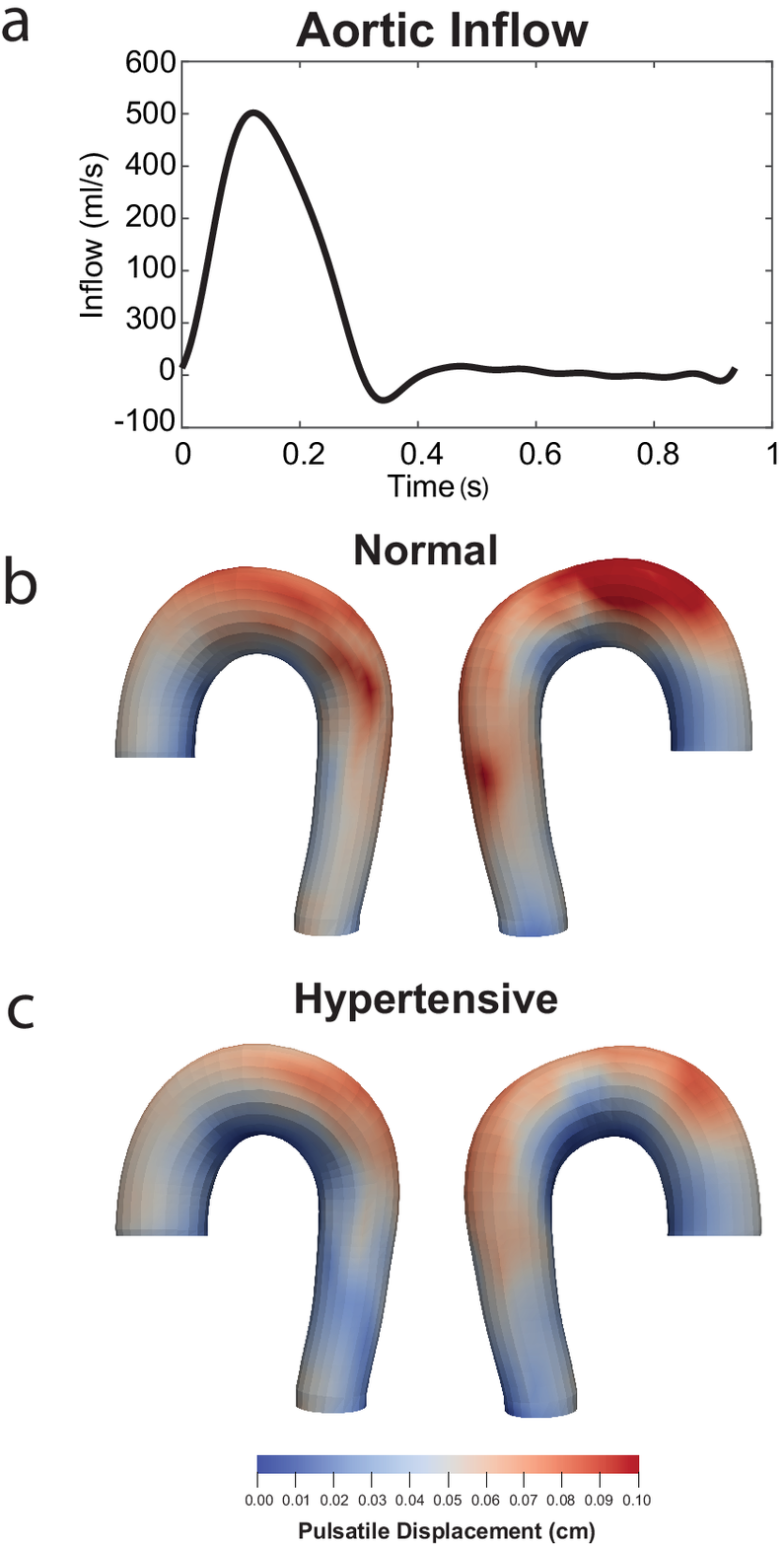}
\centering
\caption{Aortic simulations under pulsatile inflow (a). The differential remodeling under normal (b) and hypertensive (c) results in distinct responses over the cardiac cycle. Pulsatile displacement is calculated respect to the diastolic configuration.}
\label{fig:pulsatile}
\end{figure}

One advantageous feature of using the linearized stress response, as outlined in the solid-growth coupling section, is that the computational cost of pulsatile simulations is significantly reduced. Unlike the full stress formulation that requires evaluating heredity integrals at each configuration in the cardiac cycle, the linearized approach simplifies the stress calculation by considering it as solely a function of the current configuration. This computational efficiency is particularly advantageous given that the deformations occurring over a cardiac cycle remain relatively small and can be well approximated within the linear regime.\cite{baek2007theory}

Although the illustrative examples presented in this paper do not utilize pulsatile conditions, we now demonstrate the capability of our framework to handle such conditions. Specifically, we focus on the normal and hypertensive aortic models discussed in the previous section and select the configuration corresponding to $t = 800$ days. We replace the steady inflow condition with a pulsatile waveform and conduct a monolithically coupled pulsatile simulation until cyclical convergence is achieved (Figure \ref{fig:pulsatile}a). This allows us to investigate the differential remodeling effects in blood flow patterns between the normal and hypertensive aortas over a physiological cardiac cycle, providing valuable insights into their respective behaviors.

Under normal pressure conditions, the aorta has a maximum displacement of $0.13$ cm over a cardiac cycle, approximately 5\% of the aortic radius (Figure \ref{fig:pulsatile}b). After adapting to hypertensive conditions, the maximum displacement is $0.09$ cm over a cardiac cycle, approximately 3\% of the aortic radius (Figure \ref{fig:pulsatile}c). This shows how hypertensive adaptation can change the pulsatile response of the aorta as a whole. In addition, we can resolve the aortic response at specific times during the cardiac cycle. Here, we see that the maximum velocity is reduced at peak systole in the hypertensive aorta compared to the normal aorta. We also observe that in both models, the aorta preferentially displaces along the outer arch, corresponding to the area of lowest thickness (Figure \ref{fig:ps_aorta_distribution}b).

\section{Discussion}

The CMT is a mathematical framework that has been used to model the fluid-solid-growth (FSG) of biological tissues under a myriad of clinically-relevant conditions. However, previous formulations of this theory have been limited in their ability to find time-resolved solutions for patient-specific geometries that satisfy the full forms of the governing equations in each domain.\cite{figueroa2009computational, cyron2016homogenized, latorre2018mechanobiologically} Here, we introduced an FSG formulation that overcomes these limitations and is capable of simulating G\&R  in patient-specific anatomies. This represents a major step forward in the application of CMT to the study of biological tissues. This class of FSG solver is particularly necessary in pathologies with acute, dynamic changes in tissue material properties where temporal simplifications, such as equilibration or homogenization, are inappropriate.\cite{shin2005midterm} An additional benefit of the current FSG formulation is that it utilizes a linearized form of the material response within the FSI framework to iterate along the true nonlinear behavior. This linearized form, which resembles a classical St. Venant-Kirchhoff material with a known prestress and dilational volume enforcement, is readily implemented in many traditional FSI solvers and avoids the need to create architecture for evaluating heredity integrals which is not typically standard. In addition, the linearized material response provides a good approximation for material behavior over a cardiac cycle. While the examples presented in this paper primarily evaluate hemodynamic quantities under steady flow regimes, the linearized material response is seen to be applicable in the setting of pulsatile flow conditions at any given G\&R time. This opens the door to using unsteady metrics such as oscillatory shear index, residence time, and pressure pulse as inputs into G\&R which have a known impact on vascular growth and remodeling.\cite{tang2012wall, eberth2009importance, shimogonya2009can}

We validated the performance of this new FSG solver against reduced-order formulations of axisymmetrical vessels. We then demonstrated its ability to simulate patient-specific arterial and venous geometries and highlighted the importance of considering both patient-specific geometries and flow fields under homeostatic and pathological conditions. TEVGs are an important target for this solver given that their constituent makeup and geometry changes drastically over time. Here, the ability of the FSG framework to qualitatively reproduce observed TEVG behavior \textit{in-vivo} is promising. The resulting simulations highlight how surgical placement and sizing can impact TEVG outcomes and how evolving TEVG geometry can impact native vasculature remodeling due to its disruption of local hemodynamics. Future work will focus on validating TEVG behavior in a larger patient-specific cohort.

The convergence rate of the FSG formulation could be improved in future studies by using a more sophisticated form of the material stiffness tensor. Although the deformation-dependent form used in the formulation is sufficient for the proposed use cases, approximating how instantaneous changes in strain affect changes in constituent turnover would provide a truer definition of stress derivative with respect to strain. This would require finding the derivative of an implicit function, as the stress response would be a function of a heredity integral that would rely on the stress itself, but this could be approximated on a first-order basis.

As development of cardiovascular simulation tools continues, it will be important to better understand the underlying mechanisms of vascular growth and remodeling. The stimulus functions proposed herein recapitulate observed responses to hemodynamic perturbations, but incorporating more mechanistic models, such as cell-signaling models, have the potential to provide insight to a wider breadth of cardiovascular problems.\cite{estrada2021multiscale,irons2021transcript,xu2008multiscale} The ability to simulate FSG of patient-specific models also calls for the ability to calculate or estimate patient-specific parameters, including constituent material parameters, turnover rates, and perturbation gains. Although these values are difficult to measure with precision on a patient-specific basis, uncertainty quantification could be used with this FSG framework to predict the expected range of outcomes in the absence of exact parameters.\cite{fleeter2020multilevel} Understanding how stimulus functions propagate spatially will also be an important part of future FSG development as applications to patient-specific models have the potential to yield highly heterogeneous hemodynamic fields.

\section{Conclusion}
This FSG formulation of the CMT represents a significant advance in the study of biological tissues. The ability to simulate FSG in complex geometries and patient-specific models opens new possibilities for the study of cardiovascular disease progression and mechanobiology. However, further work is needed to improve the accuracy and convergence rate of the FSG formulation and understand better the underlying mechanisms of vascular growth and remodeling.

\section{Acknowledgements}
This research was supported by AHA award 835622, U.S. NIH grant 1R01HL139796, and DoD grant W81 XWH1810518, and additional funding through Additional Ventures.

\section{Dedication}
This work is dedicated to Professor T.J.R. Hughes, a true leader in computational mechanics.

\appendix

\section{Additional fluid formulation details}
\label{app:fluid}
The weak form of the fluid governing equations can be discretized using local element-level evaluation as follows:
\begin{equation} \label{eq:ns-discrete}
\int_{\Omega^f_\iota} \mathbf{w}^h \cdot \rho \left( \dot{\mathbf{u}^h} + (\mathbf{u}^h - \hat{\mathbf{u}}^h) \cdot \nabla \mathbf{u}^h - \mathbf{f}^h \right) \dd\Omega + \int_{\Omega^f_\iota} \epsilon(\mathbf{w}^h) : \sigma^f (\mathbf{u}^h,p^h) \dd\Omega - \int_{\Gamma^f_{h_\iota}} \mathbf{w}^h \cdot \mathbf{h}^h \dd \Gamma_h - \int_{\Omega^f_\iota} q^h \nabla \cdot \mathbf{u}^h \dd\Omega = \mathbf{0},
\end{equation}
where $h$ is the local element-level evaluation. Local integration is carried out using Gauss quadrature, and global integration is performed as a sum over the total number of discrete elements with
\begin{equation}\Omega_\iota \approx \Omega^h_\iota = \bigcup^{N_el}_{e=1} \Omega_e \end{equation}
describing the global domain where $\Omega_e$ represents the domain occupied by a particular element, $e$, where $N_{el}$ is the total number of elements and
\begin{equation}
\int_{\Omega^h_\iota} \dd\Omega \approx \sum^{N_{el}}_{e=1} \int_{\Omega_e} \dd\Omega
\end{equation}
describes the integration over the whole domain where the integration on a single element domain, $\Omega_e$, is carried out with Gaussian quadrature as described by Hughes.\cite{hughes2012finite} The respective variables $\mathbf{u}$, $\mathbf{w}$, $p$, and $q$ are element-level values written as 
\begin{equation}
\mathbf{u}^h = \sum_{N_{bf}} \mathbf{u}_A N_A, \quad
\mathbf{w}^h = \sum_{N_{bf}} \mathbf{w}_A N_A, \quad
p^h = \sum_{N_{bf}} p_A N_A, \quad
q^h = \sum_{N_{bf}} q_A N_A,
\end{equation}
where $\mathbf{u}_A$ and $p_A$ are trial functions for the velocity and pressure solutions, respectively, while $\mathbf{w}_A$ and $q_A$ are the respective test functions. $N_A$ are the basis function evaluations at the same point, and $N_{bf}$ is the number of basis functions used. Both are considered time-independent in this formulation.

The right-hand-side (RHS) residual vector of the Naviar-Stokes equations is computed by inputting prospective solution values for $\mathbf{u}^h$ and $p^h$ into Equation \ref{eq:ns-discrete}. We seek to minimize the residual in order to approach an exact solution. The residuals of the continuity and linear momentum equations, respectively, are as follows:
\begin{alignat}{2}
\mathbf{R}^{cont}_A &= \int_{\Omega^f_\iota} N_A \nabla \cdot \mathbf{u}^h \dd\Omega\\
\mathbf{R}^{mom}_A &= \int_{\Omega^f_\iota} N_A \hat{\mathbf{n}}_i \cdot \rho \left( \dot{\mathbf{u}}^h + (\mathbf{u}^h - \hat{\mathbf{u}}) \cdot \nabla \mathbf{u}^h - \mathbf{f}^h \right) \dd \Omega + \int_{\Omega^f_\iota} \epsilon (N_A \hat{\mathbf{n}}_i):\sigma^f(\mathbf{u}^h, p^h) \dd \Omega - \int_{\Gamma_{h_\iota}} N_A \hat{\mathbf{n}}_i \cdot \mathbf{h}^h \dd\Gamma,
\end{alignat}
where $\hat{\mathbf{n}}_i$ represents the standard Cartesian unit vector.

In our application we utilize residual-based variational multiscale (RBVMS) stabilization. The RBVMS method defines subgrid scale trial and test functions that attempt to capture modes of the final solution that are not resolved. We denote these terms with a superscript "$'$" and define them as follows:
\begin{alignat}{2}
\mathbf{u} &= \mathbf{u}^h + \mathbf{u}^{'}\\
p &= p^h + p^{'}\\
\mathbf{w} &= \mathbf{w}^h + \mathbf{w}^{'}\\
q &= q^h + q^{'}.
\end{alignat}
To formulate the stabilizing terms in RBVMS, we start by ignoring fine scale terms in the test functions. We then use integration by parts to move the derivatives onto the test functions.  The additional modes are then given as
\begin{alignat}{2}
\mathbf{u}^{'} &=  - \frac{\tau_{SUPS}}{\rho},
\end{alignat}
where $\mathbf{r}^{mom}$ and $r^{cont}$ are the residuals of the unstabilized continuity and momentum equations. These are defined in pointwise form as
\begin{alignat}{2}
\mathbf{r}^{mom}(\mathbf{u}^h, p^h) &= \rho \left( \dot{\bold{u}} + (\bold{u} - \hat{\bold{u}})\cdot \nabla \bold{u} - \bold{f} \right) - \nabla \cdot \sigma^f(\bold{u}^h, p^h)\\
r^{cont}(\bold{u}^h) &= \nabla \cdot \bold{u}^h.
\end{alignat}
We modify the terms $R^{cont}_A$ and $R^{mom}_A$ to the forms:
\begin{alignat}{2}
\mathbf{R}^{cont}_A =& \int_{\Omega^f_\iota} N_A \nabla \cdot \mathbf{u}^h \dd\Omega + \sum^{N_{el}}_{e=1} \int_{\Omega^f_e} \tau_{SUPS} \frac{\nabla N_A}{\rho} \cdot \bold{r}^{mom}(\mathbf{u}^h,p^h) \dd\Omega\\
\mathbf{R}^{mom}_A =& \int_{\Omega^f_\iota} N_A \hat{\mathbf{n}}_i \cdot \rho \left( \dot{\mathbf{u}}^h + (\mathbf{u}^h - \hat{\mathbf{u}}) \cdot \nabla \mathbf{u}^h - \mathbf{f}^h \right) \dd \Omega + \int_{\Omega^f_\iota} \epsilon (N_A \hat{\mathbf{n}}_i):\sigma^f(\mathbf{u}^h, p^h) \dd \Omega \\ \nonumber
& - \int_{\Gamma^f_{h_\iota}} N_A \hat{\mathbf{n}}_i \cdot \mathbf{h}^h \dd \Gamma + \sum^{N_el}_{e=1} \int_{\Omega^f_e} \tau_{SUPS} (\mathbf{u}^h \cdot \nabla N_A \hat{\mathbf{n}}_i) \cdot \mathbf{r}^{mom} (\bold{u}^h, p^h) \dd \Omega \\ \nonumber
& + \sum^{N_el}_{e=1} \int_{\Omega^f_e} \rho \nu_{LSIC}(\nabla \cdot N_A \hat{\bold{n}}_i) r^{cont}(\bold{u}^h) \dd\Omega \\ \nonumber
&  - \sum^{N_el}_{e=1} \int_{\Omega^f_e} \tau_{SUPS} N_A \hat{\bold{n}}_i \cdot (\bold{r}^mom(\bold{u}^h,p^h) \cdot \nabla \bold{u}^h) \dd\Omega \\ \nonumber
& - \sum^{N_el}_{e=1} \int_{\Omega^f_e} \frac{\nabla N_A \hat{\mathbf{n}}_i}{\rho} : (\tau_{SUPS} \mathbf{r}^{mom} (\mathbf{u}^h, p^h) ) \otimes (\tau_{SUPS} \mathbf{r}^{mom}(\mathbf{u}^h, p^h)) \dd\Omega.
\end{alignat}
We use previously defined values for $\tau_{SUPS}$ and $\nu_{LSIC}$ that give the relationship
\begin{alignat}{2}
\tau_{SUPS} &= (\frac{4}{\Delta t^2} + \mathbf{u}^h \cdot \mathbf{G} \mathbf{u}^h + C_I \nu^2 \mathbf{G}:\mathbf{G})^{-\frac{1}{2}}\\
\nu_{LSIC} &= (tr(\mathbf{G}) \tau_{SUPS})^{-1}.
\end{alignat}
The variable $\mathbf{G}$ is the element metric tensor (not to be confused with the constituent-specific deposition stretch tensors in the G\&R equations), defined as
\begin{equation}
\mathbf{G} = \frac{\partial \xi^\text{T}}{\partial \mathbf{x}}  \frac{\partial \xi}{\partial \mathbf{x}},
\end{equation}
where $\xi$ represents parametric coordinates of the element. The variable $C_I$ is a scalar constant that depends on topology and order, but not on the size of the element.\cite{hughes2012finite,johnson1987convergence}

Evaluation of the left-hand-side (LHS) matrix and RHS vector depends on the specific choice of time integration scheme. We employ a generalized-$\alpha$ method for time advancement. To integrate in time, we assume a given set of values at nodal points at time $t_n$, where $n$ is our timestepping index. These values are then used to determine a new set of values at some future time, $t_{n+1}$. To create a more well-defined set of equations, we use the Newmark formulation to define
\begin{equation}
\mathbf{u}_{n+1} = \mathbf{u}_n + \Delta t ((1-\gamma)\dot{\mathbf{u}}_n + \gamma\dot{\mathbf{u}}_{n+1}),
\end{equation}
where $\Delta t$ is the fluid timestep size and $\gamma$ is a non-dimensional parameter that determines the stability and accuracy of the solution. Here we utilize the value $\gamma = 0.5$. The generalized-$\alpha$ method uses two additional parameters, $\alpha_f$ and $\alpha_m$, which also determine the stability and accuracy of the time integration. We define two intermediate times, $t_{n+\alpha_f}$ and $t_{n+\alpha_m}$. We define the following relationships
\begin{alignat}{2}
\mathbf{u}_{n + \alpha_f} &= \mathbf{u}_n + \alpha_f(\mathbf{u}_{n+1} - \mathbf{u}_n)\\
\mathbf{u}_{n + \alpha_m} &= \dot{\mathbf{u}}_n + \alpha_m (\dot{\mathbf{u}}_{n+1} - \dot{\mathbf{u}}_n).
\end{alignat}
With these relationships, we can formulate the generalized-$\alpha$ method as: Given $\dot{\mathbf{u}}_n$, $\mathbf{u}$, $p_n$, find $\dot{\mathbf{u}}_{n+1}$, $\mathbf{u}_{n+1}$, $p_{n+1}$ such that
\begin{alignat}{2}
\mathbf{R}^{cont} (\dot{\mathbf{u}}_{n+\alpha_m}, \mathbf{u}_{n+\alpha_f}, p_{n+1}) &= \mathbf{0}\\
\mathbf{R}^{mom} (\dot{\mathbf{u}}_{n+\alpha_m}, \mathbf{u}_{n+\alpha_f}, p_{n+1}) &= \mathbf{0}.
\end{alignat}
Note the relationship
\begin{equation}\mathbf{u}_{n+\alpha_f} = \mathbf{u}_n + \alpha_f \Delta t (1 - \gamma) \dot{\mathbf{u}}_n + \alpha_f \gamma \Delta t \dot{\mathbf{u}}_{n+1}\end{equation}
which allows one to rewrite the equations in terms of $\dot{\mathbf{u}}_{n+ \alpha_m}$ and $\mathbf{u}_{n + \alpha_f}$ as follows
\begin{alignat}{2}
\mathbf{R}^{cont}_A =& \int_{\Omega^f_{\iota+\alpha_f}} N_A \nabla \cdot \mathbf{u}^h_{n+\alpha_f} \dd\Omega\\
\mathbf{R}^{mom}_{A_i} =&  \int_{\Omega^f_{\iota+\alpha_f}} N_A \hat{\mathbf{n}}_i \cdot \rho\left( \dot{\mathbf{u}}^h_{n+\alpha_m}  + (\mathbf{u}^h_{n+\alpha_f} - \hat{\mathbf{u}}^h_{n+\alpha_f}) \cdot \nabla \mathbf{u}^h_{n+ \alpha_f} - \mathbf{f}^h \right) \dd\Omega \\ \nonumber
& +  \int_{\Omega^f_{\iota+\alpha_f}} \epsilon (N_a \hat{\mathbf{n}}_i):\sigma^f(\mathbf{u}^h_{n+\alpha_f}, p^h) \dd\Omega - \int_{\Gamma^f_{\iota + \alpha_f}} N_A \hat{\mathbf{n}}_i \cdot \mathbf{h}^h \dd \Gamma.
\end{alignat}
These changes are extended to the stabilization terms above. We set  $\alpha_f$ and $\alpha_m$ to $0.5$ and $0.25$, respectively. With a well-defined RHS, we solve the problem iteratively on velocity and pressure using the relationship
\begin{alignat}{2}
\frac{\partial \mathbf{R}^{mom}}{\partial \dot{\mathbf{u}}_{n+1}} \Delta \dot{\mathbf{u}}_{n+1,i} + \frac{\mathbf{R}^{mom}}{\partial p_{n+1}} \Delta p_{n+1} &= -\mathbf{R}^{mom}_{i-1}\\
\frac{\partial \mathbf{R}^{cont}}{\partial \dot{\mathbf{u}}_{n+1}} \Delta \dot{\mathbf{u}}_{n+1,i} + \frac{\mathbf{R}^{cont}}{\partial p_{n+1}} \Delta p_{n+1} &= -\mathbf{R}^{cont}_{i-1}.
\end{alignat}
Each of the partial derivatives in the system of equations represents a block in the linear system. The linear system can then be written as:
\begin{equation}
\left( \begin{matrix} 
\mathbf{K} & \mathbf{G} \\
\mathbf{D} & \mathbf{L} \\
\end{matrix} \right)
\left( \begin{matrix} 
\Delta \dot{\mathbf{u}}_{n+1} \\
\Delta p_{n+1} \\
\end{matrix} \right)
=
\left( \begin{matrix} 
-\mathbf{R}^{mom} \\
-\mathbf{R}^{cont} \\
\end{matrix}\right),
\end{equation}
where we introduce the following definitions for $\mathbf{K}$, $\mathbf{G}$, $\mathbf{D}$, and $\mathbf{L}$:
\begin{alignat}{2}
\mathbf{K} &= \frac{\partial \mathbf{R}^{mom}}{\partial \dot{\mathbf{u}}_{n+1}}\\
\mathbf{G} &= \frac{\partial \mathbf{R}^{mom}}{\partial p_{n+1}}\\
\mathbf{D} &= \frac{\partial \mathbf{R}^{cont}}{\partial \dot{\mathbf{u}}_{n+1}}\\
\mathbf{L} &= \frac{\partial \mathbf{R}^{cont}}{\partial p_{n+1}}.
\end{alignat}
Within each Newton iteration, we hold convective velocity fixed so that we can linearize the equations into the appropriate matrices. We will denote this variable as $\bar{\mathbf{u}}$. These are updated at each Newton step, meaning that the LHS matrices are reassembled at every Newton iteration. The matrices are also assembled at the relevant $\alpha$ step level. We can carry out the differentiation for these matrices and incorporate the stabilization and generalized-$\alpha$ terms to define the Galerkin form of these blocks as
\begin{alignat}{2}
\mathbf{K}_{ij}
=&\int_{\Omega^f_{\iota+\alpha_f}} \delta_{ij} \rho \alpha_m N_A N_B \delta \Omega
+\int_{\Omega^f_{\iota+\alpha_f}} \delta_{ij} \alpha_f \gamma \Delta t N_A \bar{\mathbf{u}}^h_{n+\alpha_f} \cdot (\nabla N_B) \dd\Omega 
+ \int_{\Omega^f_{\iota+\alpha_f}} \delta_{ij} \mu \alpha_f \gamma \Delta t (\nabla N_A \cdot \nabla N_B) \dd\Omega\\ \nonumber
& + \int_{\Omega^f_{\iota+\alpha_f}} \mu \alpha_f \gamma \Delta t (\nabla N_A \cdot \hat{\mathbf{n}}_j) (\nabla N_B \cdot \hat{\mathbf{n}}_i ) \dd\Omega
+ \int_{\Omega^f_{\iota+\alpha_f}} \delta_{ij} \alpha_m \tau_{SUPS} \hat{\mathbf{u}}^h_{n+\alpha_f} \cdot \nabla N_B \dd\Omega\\ \nonumber
& + \int_{\Omega^f_{\iota+\alpha_f}} \delta_{ij} \rho \alpha_f \gamma \Delta t \tau_{SUPS} (\bar{\mathbf{u}}^h_{n+\alpha_f}) + \int_{\Omega^f_{\iota+\alpha_f}} \rho \alpha_f \gamma \Delta t \nu_{LSIC} (\nabla N_A \cdot \hat{\mathbf{n}}_i) (\nabla N_B \cdot \hat{\mathbf{n}}_j) \dd\Omega\\
\mathbf{G} =& -\int_{\Omega^f_{\iota+\alpha_f}} \nabla N_A \cdot \hat{\mathbf{n}}_i N_B \dd\Omega + \int_{\Omega^f_{\iota+\alpha_f}} \tau_{SUPS}(\bar{\mathbf{u}}^h_{n+\alpha_f} \cdot \nabla N_A)(\nabla N_B \cdot \hat{\mathbf{n}}_i) \dd\Omega\\
\mathbf{D} =& \int_{\Omega^f_{\iota+\alpha_f}} \alpha_f \gamma \nabla t N_A \nabla \cdot (N_B \hat{\mathbf{n}}_i) \dd\Omega 
+ \int_{\Omega^f_{\iota+\alpha_f}} \alpha_f \gamma \nabla t \tau_{SUPS} (\nabla N_A \cdot \hat{\mathbf{n}}_j)(\bar{\mathbf{u}}^h_{n+\alpha_f} \cdot \nabla N_B) \dd\Omega \\ \nonumber
 & + \int_{\Omega^f_{\iota+\alpha_f}} \alpha_m \tau_{SUPS} \nabla N_A \cdot \hat{\mathbf{n}}_j N_B \dd\Omega\\
\mathbf{L} =& \int_{\Omega^f_{\iota+\alpha_f}} \frac{\tau_{SUPS}}{\rho} \nabla N_\alpha \cdot \nabla N_B \dd \Omega.
\end{alignat}
Additional derivation details can be found elsewhere.\cite{long2013fluid}

\section{Additional solid formulation details}
\label{app:solid}

Generalized-$\alpha$ can similarly be employed in the solid formulation for time integration. The nodal displacement and acceleration at intermediate time levels are given as:
\begin{alignat}{2}
\mathbf{d}_{n+\alpha_f} &= \mathbf{d}_n + \alpha_f(\mathbf{d}_{n+1} - \mathbf{d}_n)\\
\ddot{\mathbf{d}}_{n+\alpha_m} &= \ddot{\mathbf{d}}_n + \alpha_m(\ddot{\mathbf{d}}_{n+1} - \ddot{\mathbf{d}}_n).
\end{alignat}
We again use the Newmark formula to establish a relationship between the time derivatives of the unknown vectors as
\begin{equation}
\mathbf{d}_{n+1} = \mathbf{d}_n + \Delta \dot{\mathbf{d}}_n + \frac{\Delta t^2}{2}((1-2 \beta) \ddot{\mathbf{d}}_n + 2\beta \ddot{\mathbf{d}}_{n+1}),
\end{equation}
where $\beta$ is a stability parameter. We choose
\begin{equation}
\beta = \frac{1}{4} (1 + \alpha_m - \alpha_f)^2
\end{equation}
in order to retain second-order accuracy in time.\cite{chung1993time} We construct a structural residual vector that utilizes the generalized-$\alpha$ method
\begin{equation}
\mathbf{R}^{str}_{A_i} = \int_{\Omega^s_\iota} N_A \hat{\mathbf{n}}_i \rho \cdot \ddot{\mathbf{d}}^h_{n + \alpha_m} \dd\Omega + \int_{\Omega^s_\iota} \epsilon (N_A \hat{\mathbf{n}}) \cdot \mathbf{D} \epsilon (\mathbf{d}^h_{\alpha_f}) \dd\Omega - \int_{\Omega^s_\iota} N_A \hat{\mathbf{n}}_i \rho \cdot \mathbf{f}^h \dd\Omega - \int_{(\Gamma^s_\iota)_h} N_A \hat{\mathbf{n}}_i \cdot \mathbf{h}^h \dd \Gamma,
\end{equation}
where $\mathbf{D}$ is the second tensor representation of the fourth rank tensor $\mathbb{C}$, made possible under the minor symmetry of a physically permissible $\mathbb{C}$.

The matrix system for structural mechanics is often given as
\begin{equation}\mathbf{M}\ddot{\mathbf{d}}_{n+1} + \mathbf{K} \mathbf{d}_{n+1} - \mathbf{F} = \mathbf{0}.\end{equation}
However, use of the Newmark formula requires solving in terms of $\ddot{\mathbf{d}}_{n+1}$, so we iteratively solve the system
\begin{equation}
\frac{\partial \mathbf{R}^{str}}{\partial \ddot{\mathbf{d}}_{n+1}} \Delta \ddot{\mathbf{d}}_{n+1,i} = -\mathbf{R}^{str}_{i-1},
\end{equation}
where $i$ is an iterative index. We then represent the matrix system $\frac{\partial \mathbf{R}^{str}}{\partial \ddot{\mathbf{d}}_{n+1}}$ as $\mathbf{K}$ and we can rewrite
\begin{equation}
\mathbf{K}\Delta\ddot{\mathbf{d}}_{n+1,i} = -\mathbf{R}^{str}_{i-1}.
\end{equation}
We differentiate the residual vector with respect to the unknown variable which reduces to
\begin{equation}
\mathbf{K}_{ij} = \int_{\Omega^s_\iota} \alpha_m \rho N_A N_B \delta_{ij} \dd\Omega + \int_{\Omega^s_\iota} \alpha_f \beta \Delta t^2 \hat{\mathbf{n}}_i \cdot \mathbf{B}^\text{T}_A \mathbf{D} \mathbf{B}_B \hat{\mathbf{n}}_j \dd \Omega,
\end{equation}
where $\mathbf{B}_A $ is the strain-displacement matrix for the basis function $N_A$ which satisfies
\begin{equation}
\mathbf{B}_A = \begin{bmatrix}
    N_{A,1} & 0 & 0 \\
    0 & N_{A,2} & 0 \\
    0 & 0 & N_{A,3} \\
    0 & N_{A,3} & N_{A,2} \\
    N_{A,3} & 0 & N_{A,1} \\
    N_{A,2} & N_{A,3} & 0 \\
\end{bmatrix}.
\end{equation}

\pagebreak

\bibliographystyle{unsrt} 
\bibliography{references}  

\end{document}